\documentclass{SBCbookchapter}
\usepackage[ansinew]{inputenc}
\usepackage[T1]{fontenc}
\usepackage[brazilian,english]{babel}
\usepackage{graphicx}
\usepackage{amsmath}
\usepackage{amssymb}
\usepackage{imakeidx}
\usepackage{braket}
\usepackage{authblk}
\usepackage{bbm}
\makeindex

\author[1]{Ant\^{o}nio J. G. Abel\'{e}m}
\author[2]{Gayane Vardoyan}
\author[2]{Don Towsley}
\affil[1]{Universidade Federal do Par\'{a} - UFPA \newline Faculdade de Computa\c{c}\~{a}o -- Instituto de Ci\^{e}ncias Exatas e Naturais \newline Av. Augusto Corr\^{e}a, 01 -- Guam\'{a} -- Bel\'{e}m -- PA -- 66075-110 \newline}
\affil[2]{University of Massachusetts - UMass, Amherst \newline College of Information and Computer Sciences \newline A327, Lederle Graduate Research Center, Amherst, MA, 01003 }

\title{Quantum Internet: The Future of \nobreak{Internetworking}}

\begin{document}
\setcounter{chapter}{2}
\setcounter{equation}{0}

\maketitle

\begin{abstract}
Quantum information, computation and communication, will have a great impact on our world. One important subfield will be quantum networking and the quantum Internet. The purpose of a quantum Internet is to enable applications that are fundamentally out of reach for the classical Internet. Quantum networks enable new capabilities to communication systems. This allows the parties to generate long distance quantum entanglement, which serves a number of tasks including the generation of multiparty shared secrets whose security relies only on the laws of physics, distributed quantum computing, improved sensing, quantum computing on encrypted data, and secure private-bid auctions. However, quantum signals are fragile, and, in general, cannot be copied or amplified. In order to enable widespread use and application development, it is essential to develop methods that allow quantum protocols to connect to the underlying hardware implementation transparently and to make fast and reactive decisions for generating entanglement in the network to mitigate limited qubit lifetimes. Architectures for large-scale quantum internetworking are in development, paralleling theoretical and experimental work on physical layers and low-level error management and connection technologies. This chapter aims to present the main concepts, challenges, and opportunities for research in quantum information, quantum computing and quantum networking. 
\end{abstract}
\newpage

\section{Introduction and Overview} \label{intro}
Quantum networks are a critical and highly anticipated component of an ecosystem with a broad spectrum of quantum technologies. This ecosystem will have extraordinary capabilities to effectively solve complex problems in computational sciences, communications, artificial intelligence, and data processing, and will provide a powerful capability for researchers in almost every scientific discipline \cite{Quantum_init2020}.

Connecting people or things such as computers, sensors, actuators, or databases that are in separate locations, for technical, economic, political, logistical or sometimes purely historical reasons is the main motivation for building networks, both quantum and classical. What differs is the type of data and operations involved. Quantum computers and quantum networks use quantum information rather than classical information. The analogue of the classical bit is the quantum bit, or qubit for short. Like a classical bit, a qubit has two states, but unlike a classical bit, a qubit may be in a weighted superposition of the two states, allowing certain functions to be evaluated for both input values at the same time \cite{nielsen2010quantum}.

Quantum communication is a way to transmit signals (either quantum or classical) over distances using the principles of quantum mechanics. Such signals could be used for tasks ranging from cryptography to large-scale distributed quantum computation \cite{giles2019comm}. Quantum communication takes advantage of the laws of quantum physics to protect data and can do this either through the transfer of a quantum state (entangled or not), the creation of an entangled state, or the use of a previously established entangled state. Stephen Wiesner, followed by Charles Bennett and Giles Brassard were the pioneers of modern work on quantum communication \cite{Wiesner1983}. They explained how to transmit two classical bits of information, while only transmitting one quantum bit from sender to receiver, a result dubbed superdense coding.  Bennett and Brassard's proposal \cite{bennett1984QKD} utilized the new low-level quantum capability of eavesdropping detection to create shared, secret random numbers for keying of classical cryptographic systems. However, quantum key distribution (QKD) in its basic form is limited in distance to a few hundred kilometers in optical fiber or perhaps more through free space using satellites, and is a single-application system.

One important mechanism for transmitting quantum information from one party to another across a geographic distance is teleportation, proposed first by Bennett et al. \cite{bennett1993teleporting}. Teleportation is the process of using a preshared entangled state between two parties, along with classical communication, to transport a single qubit from one party to another. This process enables a large range of distributed quantum applications. Since teleportation is executed with the help of classical communication, which proceeds no faster than the speed of light, and thus cannot be used for faster-than-light transport or communication of classical information.

Quantum communication consists of either the exchange of quantum information or the sharing of entangled quantum state between two or more parties. This allows the parties to generate long distance quantum entanglement, which serves a number of tasks including the generation of multiparty shared secrets, distributed quantum computing, improved sensing, blind quantum computing, and secure private-bid auctions. However, quantum signals are very fragile, and cannot be copied or amplified. Solutions to these problems are both similar to and different from those for classical networks \cite{VanMeter2014quantum}. According to Van Meter, all important behaviors of quantum networks arise from dealing with noise and loss using purification and quantum error correction. 

One of the major challenges of quantum communication systems lies in the transmission of quantum information with high rates over long distances in the presence of unavoidable losses \cite{Dur2007}. A solution to this is the introduction of quantum repeaters \cite{abruzzo2013quantum}, which are an interesting area of research in both experiment and theory. Quantum repeaters enable one to create entangled state between the end points of the network by first dividing the network into segments, creating entanglement across the segments, and then, connecting those entanglements to create the required long range entanglement. In other words, instead of distributing entanglement across a long link, entanglement is generated through smaller links. A combination of entanglement swapping \cite{Zukowski1993} and entanglement purification \cite{Deutsch1996} performed at each quantum repeater enables the extension of entanglement across the entire path.

In  general, the exchange of data over long distances, in topologically complex networks built on heterogeneous technologies and managed by many independent organizations, requires taking care of noise and loss. In order to enable widespread use and application development, it is essential to develop methods that allow quantum protocols to connect to the underlying hardware implementation transparently and to make fast and reactive decisions for generating entanglement in the network to mitigate limited qubit lifetimes. This can be achieved by a series of layered protocols to provide an abstraction that ultimately allows application protocols to exchange data between two end nodes without having to know any details on how this connection is actually realized. However, only preliminary functional allocation of a quantum network stack has been proposed, and just first versions of physical and link layer protocols have been developed \cite{sigcom2019} \cite{VanMeter2014quantum}.

Quantum Internet has been proposed as the key strategy to significantly scale up the number of qubits for long distance communication of quantum and classical information. However, quantum computing and networking technologies are still at an early stage of research and development (R\&D). Architectures for large-scale quantum networking and internetworking are under development, paralleling theoretical and experimental work on physical layers and low-level error management and connection technologies. Exploring how to build it will create opportunities at different levels (service, components, and modules) \cite{Wehner2018vision}.

QKD is the best-known application of a quantum Internet. However, there are many other applications that bring advantages that are unattainable with a classical network, such as secure access to remote quantum computers \cite{Kimble2008}, more accurate clock synchronization \cite{komar2014}, and scientific applications such as combining light from distant telescopes to improve observations  \cite{PhysRevLett.109.070503}. Other useful applications will likely be discovered in the next decade, as the development of a quantum Internet progresses.

This chapter aims to present the main concepts, challenges, and opportunities for research in quantum information, quantum computing and quantum networking. Besides this introductory section, this chapter is organized in four more sections. Section 2 provides a basic understanding of quantum phenomena, such as qubits, superposition, and entanglement. It describes how quantum data is represented and manipulated. Measurement, interference, decoherence, no-signaling and no-cloning theorems and other important concepts are also explained and exemplified. State-vector and Bloch sphere representations and their corresponding manipulations are discussed. The section concludes by presenting Bell-pairs and GHZ states. 

Section 3 gives a brief introduction to key quantum communication and quantum networking characteristics. We explain the concepts of teleportation, swapping, how quantum communication channels are implemented, and the new capabilities of quantum networks to communication systems. Differences between quantum and classical networks are also highlighted. How quantum networks deal with noise and loss using purification and quantum error correction will be also discussed. Then, we present QKD, the most important commercial application of quantum communication technology. We conclude the section presenting the groundwork to adapt Internet design principles to the development of quantum networks.

Section 4 presents the current status of the quantum Internet, highlighting the main initiatives, along with challenges, and research opportunities in this emerging area. The focus is on laying the groundwork to adapt Internet design principles towards the development of quantum networks. We discuss the key research challenges and open problems related to the design of a quantum network, which harness quantum phenomena with no counterpart in the classical reality, such as entanglement and superposition, to share quantum states among remote quantum devices.

Section 5 presents the general conclusions of the chapter, as well as a summary of the main contributions of the text. The key strategies of the main countries around the world to advance on the development of foundations for the quantum Internet and significantly scale up the number of qubits for long-distance communication are also presented and discussed.

\section{Quantum Information and Quantum Computing}
Until recently, every computer on the planet has operated under rules that Charles Babbage understood and that Alan Turing codified in the 1930s \cite{Turing1936}. Through the course of the computer revolution, all that has changed at the lowest level are the numbers: speed, amount of RAM and hard disk, number of parallel processors.

Since Turing, quantum computing is the first paradigm that is expected to change the fundamental scaling behavior of algorithms, making certain tasks feasible that had previously been exponentially hard. Of these, the most famous examples are simulating quantum physics and chemistry, and breaking much of the encryption that currently secures the Internet.

In order to understand quantum network operation and importance, we first must learn about the general principles upon which quantum computers are founded. In essence, a quantum computer is a device that takes advantage of quantum mechanical effects to perform certain computations asymptotically faster than a purely classical machine can \cite{VanMeter2014quantum}.

\subsection{Linear algebra and quantum mechanics} \label{Algebra}
First, let us review some basic concepts from linear algebra and describe the standard notation commonly used in quantum computing. Due to page limits and all the content that we will deal with in this chapter, we will not give strict definitions, limiting ourselves to some of the practical questions that the reader needs to know to understand the new concepts that will be presented. A rigorous definition of the linear algebra and quantum mechanics is found in Nielsen's book \cite{nielsen2010quantum}.

The basic objects of linear algebra are vector spaces. The vector space of most interest to us is $\mathbbmss{C}^n$, the space of all n-tuples of complex numbers, $(z_1,\;\ldots,\;z_n)$. The elements of a vector space are called vectors.

Quantum mechanics is our main motivation for studying linear algebra. Paul Dirac and Erwin Schr\"{o}dinger played important roles in quantum mechanics. Among other contributions, Dirac introduced the bra-ket notation. A column vector in Dirac's ket notation is  represented by the ket  $\ket{\psi}$. It is defined as: 
\begin{equation}
 \ket{\psi} = \left( \begin{array}{c} a_0 \\
a_1\\
\vdots \\
a_{N-1} \\
\end{array}\right) 
\end{equation}
where $a_i \in \mathbbmss{C}, \; i = 0,\; 1, \; \dots, \; N-1 $. The bra $\bra{\psi}$ is the adjoint (conjugate transpose) of $\ket{\psi}$ given as:

\begin{equation}
\bra{\psi} = \left( \begin{array}{r} a^*_0, \; a^*_1, \; \ldots, \; a^*_{N-1}
\end{array}\right)
\end{equation}
where $a^{*}$ is the complex conjugate of $a$. That is, if $ a=x+iy $, then $a^{*}=x-iy $. All vector spaces are assumed to be finite dimensional, unless otherwise noted. 

A linear operator between vector spaces $V$ and $W$ is defined to be any function $A : V \rightarrow W$ which is linear in its inputs:
\begin{equation}
A \left( \sum_{i} a_i\ket{v_i} \right) = \sum_{i} a_i\; A \left( \ket{v_i} \right)  
\end{equation}
When we say that a linear operator $A$ is defined on a vector space, $V$, we mean that $A$ is a linear operator from $V$ to $V$. An important linear operator on any vector space $V$ is the identity operator, $I_V$, defined by the equation $I_V \Ket{v} \equiv \Ket{v}$ for all vectors $\Ket{v}$. Another important linear operator is the zero operator, which we denote $0$. The zero operator maps all vectors to the zero vector, $0\Ket{v} \equiv 0$.

An inner product is a function that takes as input two vectors $\ket{v}$ and $\ket{w}$ from a vector space and produces a complex number as output. It can be written as $(\ket{v}, \ket{w})$,  or as $\langle v | w \rangle$. The first notation is commonly used in linear algebra, while the second is the standard quantum mechanical notation. So, we can write the inner product of $\Ket{\psi_A}$ and $\Ket{\psi_B}$ as:
\begin{equation}
(\ket{\psi_A}, \ket{\psi_B}) =  \langle \psi_A|\psi_B \rangle = \sum_{i=0}^{N-1} a^{*}_i b_i    
\end{equation}
It is important to note that, in the finite dimensional complex vector spaces that come up in quantum computation and quantum information, a Hilbert space \footnote{A Hilbert space is an abstract vector space possessing the structure of an inner product that allows length and angle to be measured.} is exactly the same thing as an inner product space, that is a vector space $V$ with an inner product on $V$  \cite{linear_alg1997}.

An eigenvector of a linear operator $A$ on a vector space is a non-zero vector $\ket{v}$ such that $A\ket{v} = \lambda \ket{v}$, where $\lambda$ is a complex number known as the eigenvalue of A corresponding to $\ket{v}$.

Suppose $A$ is any linear operator on a Hilbert space, $V$. It turns out that there exists a unique linear operator $A^\dagger$ on $V$ such that for all vectors $\ket{v}, \ket{w} \in V$,
\begin{equation}
\left( \ket{v},\; A \ket{w} \right) = \left(A^\dagger \ket{v},\; \ket{w} \right)  \label{Adjoint}
\end{equation}
This linear operator is known as the adjoint or Hermitian conjugate of the operator A. From Equation (\ref{Adjoint}) , we can obtain that $(AB)^\dagger = B^\dagger A^\dagger$. By convention, if $\ket{v}$ is a vector, then we define $\ket{v}^\dagger \equiv \bra{v}$. With this definition we also see that $\left(A\ket{v}\right)^\dagger \equiv \bra{v} A^\dagger$. An operator A whose adjoint is A is known as a Hermitian or self-adjoint operator.  

A special subclass of Hermitian operators is extremely important, the positive operators. A positive operator on a complex Hilbert space is necessarily a symmetric operator and has a self-adjoint extension that is also a positive operator. A positive operator $A$ is one for which the inner product between $\bra{\psi}$ and $A \ket{\psi}$ is greater or equal to $0$ (i.e. $\bra{\psi}A\ket{\psi} \geq 0$) for all $\ket{\psi}$. A positive definite operator $A$ is one for which $\bra{\psi}A\ket{\psi} > 0$ for all $\ket{\psi} \not = 0$. 

The notation $U$ will generically be used to denote a unitary operator or matrix. A matrix $U$ is said to be unitary if $U^\dagger U = I$. Similarly an operator $U$ is unitary if $U^\dagger U = I$. So, an operator is unitary if and only if each of its matrix representations is unitary. A unitary operator also satisfies $UU^\dagger = I$. 

The trace of a matrix is an important matrix function, very useful in quantum mechanics. The trace of $A$ is the sum of its diagonal elements,

\begin{equation}
tr(A) \equiv \sum_{i} A_{ii}    
\end{equation}

The trace is cyclic, $tr(AB) = tr(BA)$, and linear, $\;tr(A + B) = tr(A) + tr(B)$, \;$tr(zA) = z tr(A)$, where $A$ and $B$ are arbitrary matrices, and $z$ is a complex number. The trace of an operator $A$ is the trace of any matrix representation of $A$ \cite{nielsen2010quantum}.

An alternate formulation to describe quantum mechanics is to use a tool known as the density operator or density matrix. This alternate formulation is mathematically equivalent to the state vector approach, but it provides a more convenient language for thinking about some commonly encountered scenarios in quantum mechanics. The density operator is used to describe quantum systems whose state is not completely known. 
Suppose a quantum system is in one of a number of states $\ket{\psi_i}$, where $i$ is an index, with respective probabilities $p_i$. $\{{p_i, \ket{\psi_i}}\}$ is usually called an ensemble of pure states. The density operator for the system is  formally  defined  as  the  outer  product  of  the  $\ket{\psi_i}$  and  its  conjugate $\bra{\psi_i}$. 

\begin{equation}
\rho \equiv \sum_{i} p_i\; \ket{\psi_i}\; \bra{\psi_i}.    
\end{equation}
The density operator is often known as the density matrix. An operator $\rho$ is the density operator associated with some ensemble $\{{p_i, \ket{\psi_i}}\}$ if and only if it satisfies the conditions:

\begin{enumerate}
    \item Trace condition: $\rho$ has trace equal to one. 
    \item Positivity condition: $\rho$ is a positive operator.
\end{enumerate}
The proof of this theorem and a detailed discussion about density operator is provided by Nielsen et al. \cite{nielsen2010quantum}.

The notation we review in this section is summarized in Table \ref{notation}:
\begin{table}[h!]
\caption{Summary of notations}
\begin{center}
\begin{tabular}{ c | l }
\hline
Notation & Description \\
\hline
$z^*$ & Complex conjugate of the complex number $z$. \\
$\ket{\psi}$ & Vector, known as a ket. \\
$\bra{\psi}$ & Vector complex conjugate to $\ket{\psi}$, known as a bra. \\
$\langle \varphi|\psi \rangle$ & Inner product between the vectors $\ket{\varphi}$ and $\ket{\psi}$. \\
$A^\dagger$ & Hermitian transpose or adjoint of the $A$ matrix.\\
$\bra{\varphi}A\ket{\psi}$ & Inner product between $\ket{\varphi}$ and $A\ket{\psi}$. \\
$tr(A)$ & Trace of a matrix. \\
$\rho$ & Density operator or density matrix. \\
\hline
\end{tabular}
\label{notation}
\end{center}
\end{table}

\subsection{Quantum bits: qubits} \label{q_bits}
The fundamental concept of classical computation and classical information is the bit. Quantum computation and quantum information are built on an analogous concept, the quantum bit, or qubit. While a classical bit is a data element with two values, 0 and 1, a qubit is represented using either as a true two-level system, such as the polarization of a photon or the spin of an electron, or a pseudo-two-level system, such as two energy levels of an atom that can be treated as a two-level system \cite{nielsen2010quantum}. 

The difference between a classical bit and a qubit is that a qubit can be in a superposition of the two states. The state of a qubit can be written as:
\begin{equation}
    \ket{\psi} = \alpha \ket{0} + \beta \ket{1}
\label{qubitstate}    
\end{equation}
where $\alpha$ and $\beta \in \mathbbmss{C}$ . Put another way, the state of a qubit is a vector in a two-dimensional complex vector space. The special states $\ket{0}$ and $\ket{1}$ are known as computational basis states, and form an orthonormal basis for this vector space.

A classical bit may be examined many times to determine whether it is in the state 0 or 1. However, a qubit generally cannot be examined to determine its full quantum state, that is, the values of $\alpha$ and $\beta$. Instead, quantum mechanics tells us that we can only acquire much more restricted information about the quantum state through a measurement operation. When we measure a qubit we get either the result 0, with probability $|\alpha|^2 $, or the result 1, with probability $|\beta|^2$. Naturally, $|\alpha|^2 + |\beta|^2 = 1$, since the probabilities must sum to one. Geometrically, we can interpret this as the condition that the qubit's state be normalized to length 1. Thus, in general a qubit's state is a unit vector in a two-dimensional complex vector space.

\begin{figure}[h!t]
\centering
\includegraphics[width=8cm, height=9cm]{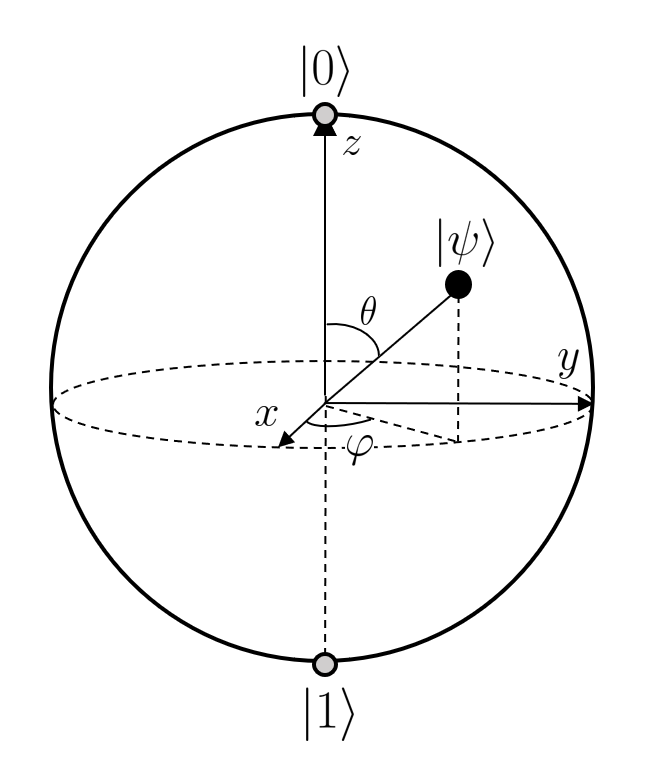}
\caption{Bloch sphere}
\label{blochsphere}
\end{figure}

As $|\alpha|^2 + |\beta|^2 = 1$,  $\ket{\psi}$ can also be expressed as:
\begin{equation}
    \ket{\psi} = \cos {\frac{\theta}{2}}\ket{0} + e^{i\varphi} \sin {{\frac{\theta}{2}} \ket{1}}.
\end{equation}
This equation facilitates the representation of a qubit's state in a three-dimensional sphere, called {\it Bloch sphere}, as shown in Figure ~\ref{blochsphere}. The numbers $\theta$ and $\varphi$ define a point on the unit three-dimensional sphere. The south-north axis is the Z-axis, the positive X-axis is toward the reader (out of the page or screen) and the Y-axis is right-left.

The {\it Bloch sphere} provides a useful means of visualizing the state of a single qubit, and often serves as an excellent testbed for ideas about quantum computation and quantum information. If the vector points at the north pole, our qubit is in the $\ket{0}$ state, and if it
points at the south pole, the qubit is in the $\ket{1}$ state. When the unit vector points toward you, that is the $(\ket{0}+\ket{1})/\sqrt{2}$ state; when it points away from you, that is the $(\ket{0}-\ket{1})/\sqrt{2}$ state. These two states are called the $\ket{+}$and $\ket{-}$ (read "ket plus" and "ket minus") states. The positive Y-axis is $(\ket{0}+i\ket{1})/\sqrt{2}$ and the negative Y-axis is $(\ket{0}-i\ket{1})/\sqrt{2}$. The phase is the position of the vector about the Z-axis. 

\subsection{Multiple qubits} \label{multi_qbits}
While the state vector is two-dimensional for a single qubit, the state vector is $N = 2^{n}$ dimensional for a n-qubit register. Suppose we have a two qubit system, it will have four computational basis states denoted $\ket{00},\ket{01},\ket{10},\ket{11}$. A related set of two or more qubits is commonly referred to as a quantum register \cite{Hagouel2012}.

A pair of qubits can also be in superpositions of these four states.  In this way, the state vector describing the two qubits is

\begin{equation}
    \ket{\psi} = \alpha_{00} \ket{00} + \alpha_{01} \ket{01} + \alpha_{10} \ket{10} + \alpha_{11} \ket{11},  
\label{mqubitstate}    
\end{equation}
where each computational basis state is associated with a complex coefficient, called an amplitude. 

An important two qubit state is the Bell state or EPR pair

\begin{equation} \label{BP}
    (\ket{00} + \ket{11})/{\sqrt{2}}.
\end{equation}

It is the key ingredient in quantum teleportation, which forms the foundation of much of quantum networking, as we will see in Section \ref{q_comm_net}. 

Bell pairs have the property that the qubits are correlated or entangled. Quantum entanglement is a quantum mechanical phenomenon in which the quantum states of two or more objects have to be described with reference to each other, even though the individual objects may be spatially separated. These correlations have been the subject of intense interest ever since a famous paper by Einstein, Podolsky, and Rosen \cite{Einstein1935}. In the 1960s, John Bell extending and clarifying the work of Einstein et al. and proved that the measurement correlations in the Bell state are stronger than could ever exist between classical systems \cite{bell_aspect_2004}.

If we generalize and consider a system of n qubits, the computational basis states of this system are of the form $\ket{x_1\;x_2\;x_3\;\ldots\;x_n}$,  and a quantum state of such a system is specified by $2^n$ amplitudes. We use the tensor product to compose the state of two or more qubits into one vector, or operations on multiple qubits into a single operator \cite{Bourbaki1989}.% Suppose $P$ and $Q$ are vector spaces of dimension $m$ and $n$ respectively. For convenience we also suppose that $P$ and $Q$ are Hilbert spaces. Then $P \otimes Q$ is an $m\;n$ dimensional vector space. The elements of $P \otimes Q$ are linear combinations of tensor products $\ket{p} \otimes \ket{q}$ of elements $\ket{p}$ of $P$ and $\ket{q}$ of $Q$ .

\subsection{Quantum gates and quantum circuits} \label{Q_gates}
Quantum computation proceeds by taking a set of qubits, modifying their states such that a "computation" of some interest is performed and reading out the result so that we learn what happened. Analogous to the way a classical computer is built from an electrical circuit containing wires and logic gates, a quantum computer is built from a quantum circuit containing wires and elementary quantum gates to carry around and manipulate the quantum information \cite{nielsen2010quantum}.

In the circuit model, quantum computations are decomposed into separate gates and can be organized more or less along the lines of classical circuits. In order for our computational capabilities to be "universal", we must be able to reach any point on the Bloch sphere for a single qubit.

Consider, for example, classical single bit $NOT$ gate $(X)$, whose operation is defined by its truth table, in which $0 \rightarrow 1$ and $1 \rightarrow 0$, that is, the 0 and 1 states are interchanged. In the quantum world, a single-qubit operation can be any rotation on the Bloch sphere. Rotations about the axes of the Bloch sphere can be described in terms of the Pauli matrices, which are a set of three 2 x 2 complex matrices which are Hermitian and unitary, that arise in Pauli's treatment of spin in quantum mechanics \cite{nielsen2010quantum}. However, specifying the action of the gate on the states $\ket{0}$ and $\ket{1}$ does not tell us what happens to superpositions of the states $\Ket{0}$ and $\ket{1}$, without further knowledge about the properties of quantum gates. In fact, the quantum $NOT$ gate acts linearly, that is, if $\ket{\psi} = \alpha \ket{0} + \beta \ket{1}$, then $X\ket{\psi} = \alpha \ket{1} + \beta \ket{0}$

Unlike the classical case where there is only one nontrivial gate ($NOT$ gate), there are many non-trivial single-qubit gates. Two important ones are the Z gate and the Hadamard gate $(H)$. The first one leaves $\ket{0}$ unchanged, and flips the sign of $\ket{1}$ to give $-\ket{1}$.

\begin{equation}
    Z \equiv \left[
    \begin{array}{cc}
    1 & 0 \\
    0 & -1 \\
    \end{array}
    \right]
\end{equation}
The second one, the Hadamard gate has representation:
\begin{equation}
    H \equiv \frac{1}{\sqrt{2}} \left[
    \begin{array}{cc}
    1 & 1 \\
    1 & -1 \\
    \end{array}
    \right]
\end{equation}
when applied to $\ket{0}$, it returns $H\ket{0}=(\ket{0} + \ket{1})/\sqrt{2}$, while when applied to $\ket{1}$, it returns $H\ket{1}=(\ket{0} - \ket{1})/\sqrt{2}$.

\begin{figure}[h!]
\centering
\includegraphics[width=11cm, height=4cm]{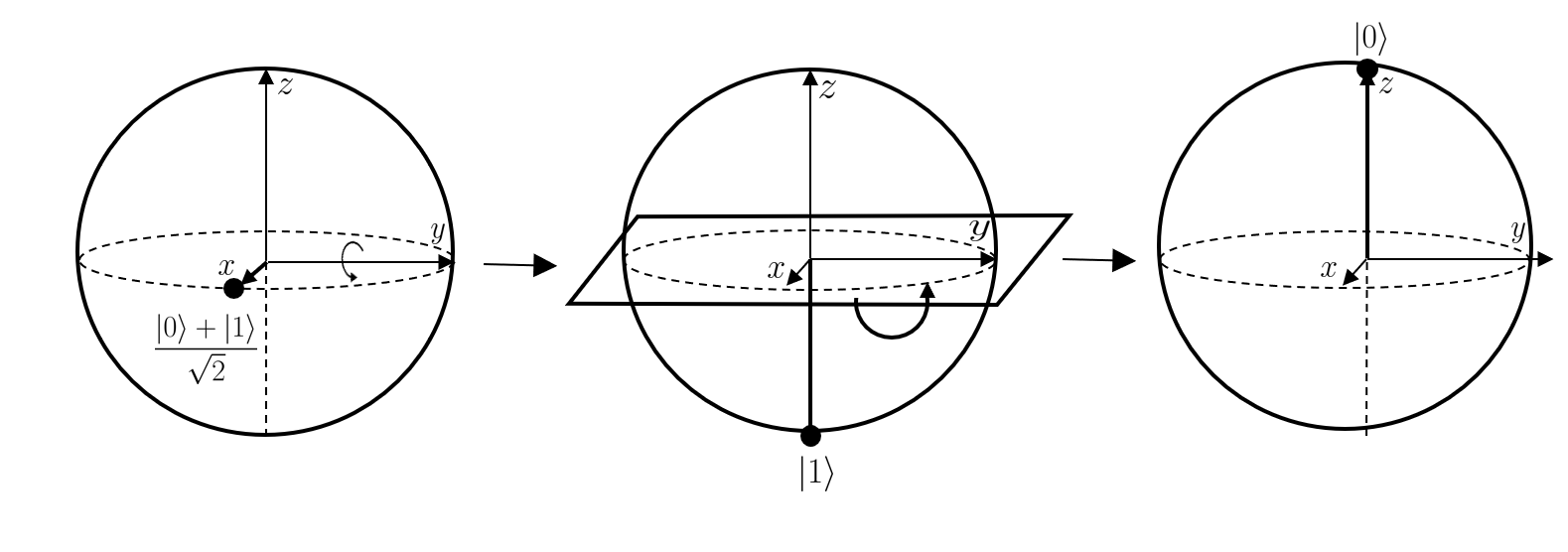}
\caption{ Hadamard gate on the Bloch sphere, acting on the input state $(\ket{0} + \ket{1})/\sqrt{2}$}
\label{hadasphere}
\end{figure}

\begin{figure}[h!]
%\centering
\begin{flushright}
\includegraphics[]{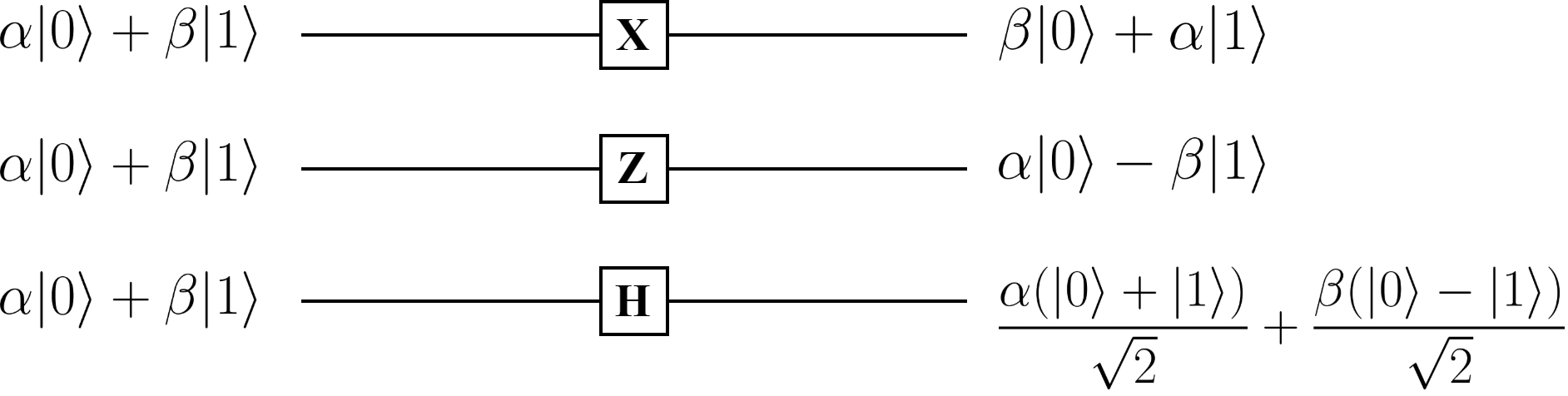}
\end{flushright}
\caption{More important single qubit gates}
\label{singlegates}

\end{figure}

The Hadamard gate is one of the most useful quantum gates and is worth visualizing its operation on the Bloch sphere as illustrated in Figure \ref{hadasphere}. Geometrically, we visualize the Hadamard operation as a $90^o$ rotation about the Y-axis, followed by a $180^o$ rotation about the X-axis \cite{nielsen2010quantum}. Figure \ref{singlegates} summarizes single qubit gates presented.

As computations involve more than one qubit, let us generalize from one to multiple qubits. First, consider the controlled-NOT gate, or CNOT. This gate has two input qubits, known as the control qubit and the target qubit, respectively. If the control qubit is one, a NOT operation is performed on the target qubit; if the control qubit is zero, the target bit is left unchanged. The output is the exclusive OR (XOR) of the two qubits, and one of the input qubit and may be summarized as $\Ket{A, B} \rightarrow \ket{A, A \oplus B}$. Table \ref{truthtable} shows the truth table for a CNOT with A as the control bit and B as the target bit.

\begin{table}[h!]
\caption{Controlled-NOT truth table}
\begin{center}
\begin{tabular}{ c | c }
\hline
Input & Output \\
\hline
AB & AB \\
$\ket{00}$ & $\ket{00}$ \\
$\ket{01}$ & $\Ket{01}$ \\
$\ket{10}$ & $\ket{11}$ \\
$\ket{11}$ & $\ket{10}$ \\
%A \; B & A\; B \\
%0 \; 0 & 0\; 0 \\
%0 \; 1 & 0\; 1 \\
%1 \; 0 & 1\; 1 \\
%1 \; 1 & 1\; 0 \\
\hline
\end{tabular}
\label{truthtable}
\end{center}
\end{table}

\begin{figure}[h!t]
\centering
\includegraphics[width=12cm, height=3cm]{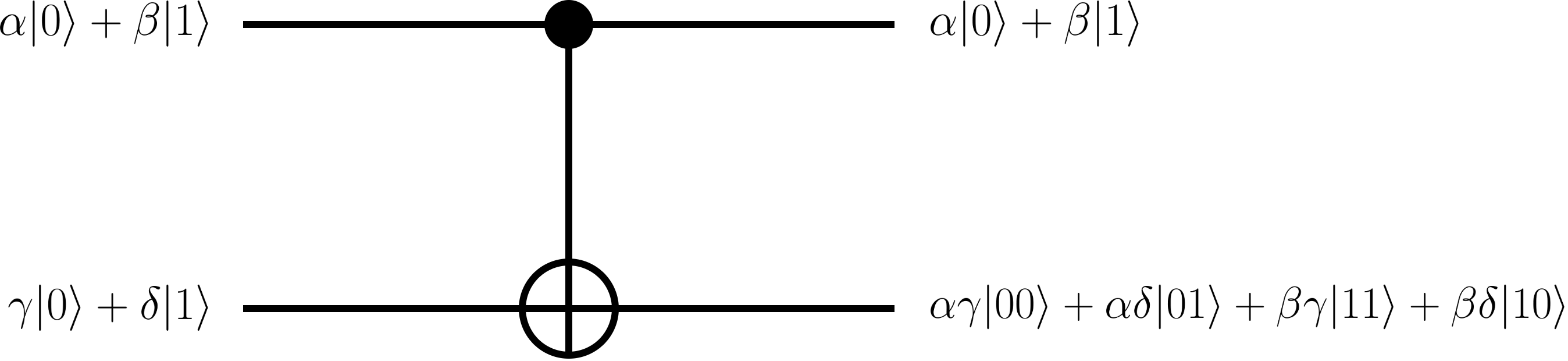}
\caption{Controlled-NOT gate}
\label{multigate}
\end{figure}

The circuit representation for the $CNOT$ is shown in the Figure \ref{multigate}. The top line represents the control qubit, while the bottom line represents the target qubit.

A quantum computation, in the abstract, is a unitary transformation on an initial quantum state, creating desired states, which we can then measure. A complete unitary transform on $n$ qubits, of course, is a $2^n\; x\; 2^n$ matrix; therefore, direct construction of the unitary to implement a complex function of more than a few qubits is difficult. A quantum circuit effects the overall transform via a series of smaller gates (generally, one to three-qubit gates) applied in a prescribed order on the appropriate qubits.

\begin{figure}[h!t]
\centering
\includegraphics[]{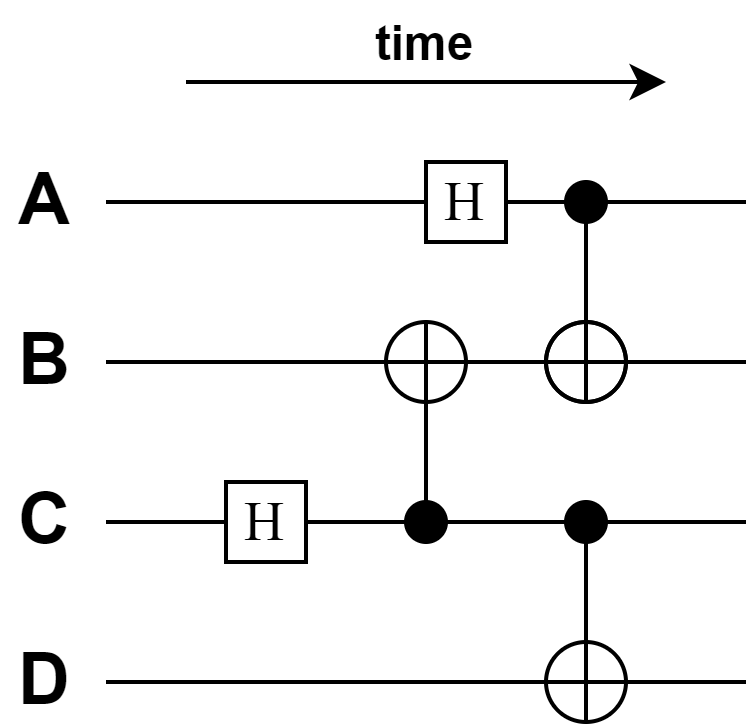}
\caption{Simple quantum circuit}
\label{simplecircuit}
\end{figure}

Researchers have found several methods for decomposing a specific unitary transform into a series of small gates or operations that we know how to implement. Figure \ref{simplecircuit} shows a simple example of a four qubit quantum circuit. This circuit consists of two Hadamard gates and three CNOT gates. Gates on different qubits can be executed at the same time, as shown by the vertical alignment. 

\subsection{Measurement} \label{measurement}
In quantum physics, measurement is the testing or manipulation of a physical system in order to yield a numerical result. The predictions that quantum physics makes are in general probabilistic \cite{q_theory2001}. 

As presented in Section \ref{q_bits}, a qubit described by the Equation (\ref{qubitstate}) can exist in a continuum of states between $\ket{0}$ and $\ket{1}$. When a qubit is measured, it only ever gives $'0'$ or $'1'$ as the measurement result - probabilistically. The measurement changes the state of a qubit, collapsing it from its superposition of $\ket{0}$ and $\ket{1}$ to the specific state consistent with the measurement result. For example, consider that a qubit is in the state $\ket{+}$

\begin{equation} \label{ket+}
    (\ket{0} + \ket{1})/{\sqrt{2}}.
\end{equation}
If measurement gives $0$, then the post-measurement state of the qubit will be $\ket{0}$.

In an analogous way it is possible in principle to measure a quantum system of many qubits with respect to an arbitrary orthonormal basis. For two or more qubits, we can measure either the entire system or only part. Measuring a single qubit can alter the state of the system. For example, consider the state vector of the Equation (\ref{mqubitstate}), describing the two qubit in superpositions. The measurement result $x (= 00, \; 01,\; 10 \; or \; 11)$ occurs with probability $|\alpha_x|^2$, with the state of the qubits after the measurement being $\ket{x}$, in a similar way to the case for a single qubit.

As we mentioned in Section \ref{multi_qbits}, the Bell state has the important property that the measurement outcomes are entangled. Upon measuring the first qubit, one obtains two possible results: $0$ with probability $1/2$, leaving the post-measurement state $= \ket{00}$, and $1$ with probability $1/2$, leaving the post-measurement state $=\ket{11}$. As a result, a measurement of the second qubit always gives the same result as the measurement of the first qubit. In others words, measuring one qubit has determined the state of the other. 

\begin{figure}[h!t]
\centering
\includegraphics[width=8cm, height=4cm]{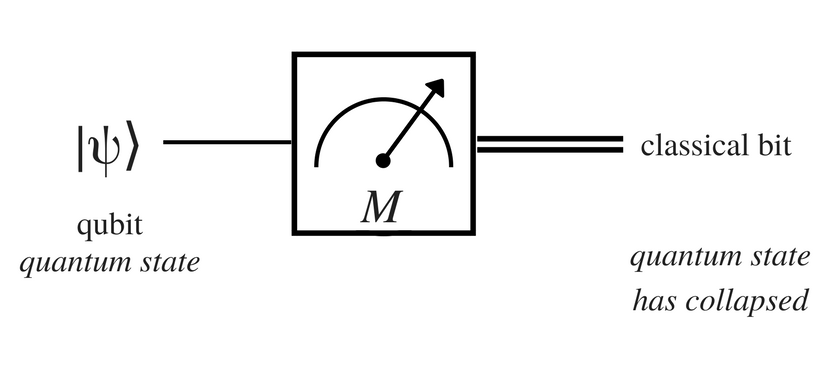}
\caption{symbol for measurement.}
\label{measurement_symbol}
\end{figure}

In the literature, the measurement operation is represented by a 'meter' symbol, as shown in Figure \ref{measurement_symbol}. The input is a qubit in a state $\ket{\psi}$ and the output is a classical bit, distinguished from a qubit by drawing it as a double-line wire. For readers interested in this fascinating topic, one good place to start studying is Preskill's lecture notes \cite{preskill98}.

\subsection{Interference, decoherence and fidelity} \label{Interf_decoh_fidelity}
In physics, interference is the combination of two or more waveforms to form a resultant wave, in which the displacement is either reinforced or canceled \cite{Interference_1986}. Quantum interference can happen between particles that arrive at the same position or quantum state but by different paths. Quantum interference, a byproduct of superposition, is what allows us to bias the measurement of a qubit toward a desired state or set of states. 

Decoherence is the gradual decay of the state of a system. It happens because quantum states are very fragile: excited atoms decay and spins of electrons and atomic nuclei spontaneously flip. Any quantum system is affected through interactions with its environment, leaking information about its state out into the environment where it cannot be recovered.

When decoherence occurs, measurement of the system may not produce the desired results, causing the failure of our quantum algorithm. The two key measures of decoherence are the T1 and T2 times. T1 is the energy relaxation time, and T2 is the phase relaxation time. Both processes are memoryless, with probabilistic behavior. The amount of time we can count on the state of a qubit remaining in a usable state is a function of the minimum of T1 and T2. Researchers determine these values experimentally, and an important area of device research is extending these times by careful engineering of the environment and control system.  

Fidelity is used to track the quality of the state. Fidelity ranges from 0 to 1.0, with the latter being perfect. It is, essentially, the probability that our qubit or set of qubits is actually in the state we believe it ought to be in. In other words, fidelity of a state corresponds to how imperfect it is in relation to a desired state \cite{Joz1994, Liang_2019}. It is not a metric, but has some useful properties and it can be used to define a metric on this space of density matrices. We will define the fidelity as

\begin{equation}
   F =  \bra{\psi} \rho \ket{\psi}
\label{fidelity}    
\end{equation}
where $0 \leq F \leq 1$ is the fidelity \footnote{In literature, the fidelity is often defined as $F = \sqrt{\bra{\psi} \rho \ket{\psi}}$, but in keeping with Jozsa's definition \cite{Joz1994}, adopted also in Van Meter's book \cite{VanMeter2014quantum}, we dispense with the square root.}, $\ket{\psi}$ is the state we think we have created and $\rho$ is the density matrix of the actual state. 

The fidelity can also be thought of the overlap of our actual state with the desired state. The fidelity is 1.0 for a pure state and declines as noise in the system degrades the quality of the state. Consider, for example, we are initializing a two-qubit register to the $\psi = \ket{00}$ state, but that the initialization process is imperfect. To learn how imperfect, we repeat the process a number of times and measure the state, to build up a statistical picture of our ability to create the desired state. From this process, we obtain the density matrix $\rho_\psi$ and, then, we can calculate the fidelity $F_\psi$ for our desired state. For an $n$-qubit state, the completely mixed state in which all qubits are random, we have $F = 2^{-n}$.

\subsection{Bell pairs and GHZ states} \label{entangled}
As we mentioned in Section \ref{multi_qbits}, the best known two qubit entanglement involving two parts sharing two qubits is the Bell state.
In the addition to the one listed in $(\ref{BP})$ there are three others forms of Bell pairs:

\begin{equation} \label{BP2}
    (\ket{00} - \ket{11})/{\sqrt{2}}
\end{equation}
\begin{equation}\label{BP3}
    (\ket{01} + \ket{10})/{\sqrt{2}}
\end{equation}
\begin{equation} \label{BP4}
    (\ket{01} - \ket{10})/{\sqrt{2}}
\end{equation}

With Bell pairs $(\ref{BP})$ and $(\ref{BP2})$, a measurement of one qubit will result in both qubits being zero or both qubits being one, with equal probability. For example, Alice may hold one qubit, while Bob holds another one, at an arbitrary distance apart without the behavior of the Bell pair changing. When Alice measures her qubit and finds a one, she will be sure that when Bob measures his qubit, it will be a one. Likewise, if she measures zero, Bob will measure zero. In $(\ref{BP3})$ and $(\ref{BP4})$, in contrast, if Alice measures a one, Bob will measure a zero and vice-versa. Moreover, this effect does not change if Bob measures his qubit first or if they both measure their qubits simultaneously.

There are other, larger multi-party entangled states that are useful for a variety of tasks, the Greenberger-Horne-Zeilinger (GHZ) state \cite{Bravyi_2006}. It was first studied by Daniel Greenberger, Michael Horne and Anton Zeilinger in 1989. It involves at least three subsystems (particle states, or qubits) and allows to observe extremely non-classical properties of the state \cite{greenberger2007going}.

The GHZ state is an entangled quantum state of $M > 2$ subsystems. In simple words, it is a quantum superposition of all subsystems being in state 0 or all of them being in state 1, represented by:

\begin{equation} \label{GHZ}
    \frac{\ket{000...} + \ket{111...}}{\sqrt{2}}
\end{equation} 

There is no standard measure of multi-partite entanglement because different, not mutually convertible, types of multi-partite entanglement exist. Nonetheless, many measures define the GHZ state to be maximally entangled state. 

GHZ states are used in several protocols in quantum communication and cryptography, including any of several common forms of three-party or larger states, for example, in secret sharing or in the Quantum Byzantine Agreement \cite{VanMeter2014quantum}. 

\section{Quantum Communication and Quantum Networks} \label{q_comm_net}
Quantum communication consists of either the exchange of quantum information or the sharing of entangled quantum state between two or more parties. It takes advantage of the laws of quantum physics to protect data and offers new functionality over classical communication. Its most interesting application is protecting information channels against eavesdropping by means of quantum cryptography. %\cite{awschalom2019development}.

To transport qubits from one node to another, we need communication Channels. Quantum communication channels are implemented by sending states of light down a physical channel. These states may be single photons or other quantum optical states with either large or small numbers of photons. For the purpose of quantum communication, standard telecom fibers can be used or free space.  It may involve a single transmitter and receiver, or multiple receivers that can individually be enabled or disabled in a shared bus configuration. A link uses a quantum channel and associated classical channel to connect two or more nodes.

To make maximum use of communication infrastructure, we also require optical switches capable of delivering qubits to the intended quantum processor. These switches need to preserve quantum coherence, which make them more challenging to realize than standard optical switches. Current commercial switches have various problems that make them unsuitable for rerouting entangled photons. Those that are made of micro-electromechanical components keep entangled states intact but operate too slowly. Other optoelectronic switches either add too much noise so that single photons are difficult to detect, or they completely destroy the quantum information. The utilization of a quantum switch provides significant advantages for a number of problems, ranging from quantum computation and quantum information processing, through non-local games to quantum communication \cite{Caleffi_2020}. 

Quantum networks enable the secure transmission and exchange of quantum communications between distinct, physically separated quantum processors, or endpoints. However, quantum signals are weak and very fragile and cannot be copied or amplified. Consequently, quantum operations are needed to exchange data over long distances to deal with noise and loss.

This section provides a brief introduction to quantum communication and networking. We present the concept of teleportation, swapping, and how quantum communication channels are implemented. Differences between quantum and classical networks are also highlighted. How quantum networks deal with noise and loss using purification and quantum error correction will be also discussed. Then, we present the most important commercial application of quantum communication technology, Quantum Key Distribution (QKD). We conclude the section presenting the groundwork to adapt Internet design principles to the development of quantum networks.

\subsection{Quantum teleportation} \label{telep}
As mentioned in Section \ref{intro}, quantum teleportation is the process by which quantum information can be transmitted from one location to another, with the help of classical communication and previously shared quantum entanglement between two parties. 

Teleportation is best described through the communication between Alice and Bob. Suppose Alice and Bob generated a Bell pair, each taking one qubit of the Bell pair. Alice must deliver a qubit $\ket{\psi}$ to Bob. She does not know the state of the qubit, and can only send classical information to Bob. Alice interacts the qubit $\ket{\psi}$ with her half of the EPR pair, and then measures the two qubits in her possession, obtaining one of four possible classical results, 00, 01, 10, and 11. She sends this information to Bob. Depending on Alice's classical message, Bob performs one of four operations on his half of the EPR pair. By doing this he can recover the original state $\ket{\psi}$.

Teleportation consumes exactly one Bell pair. Two classical bits of measurement result must be communicated. Hence, the time to execute one operation corresponds to the time needed to transmit the classical information. Careful engineering may allow pipelining of this operation with others \cite{bennett1993teleporting}.

\begin{figure}[h!t]
\centering
\includegraphics[width=14cm, height=5.25cm]{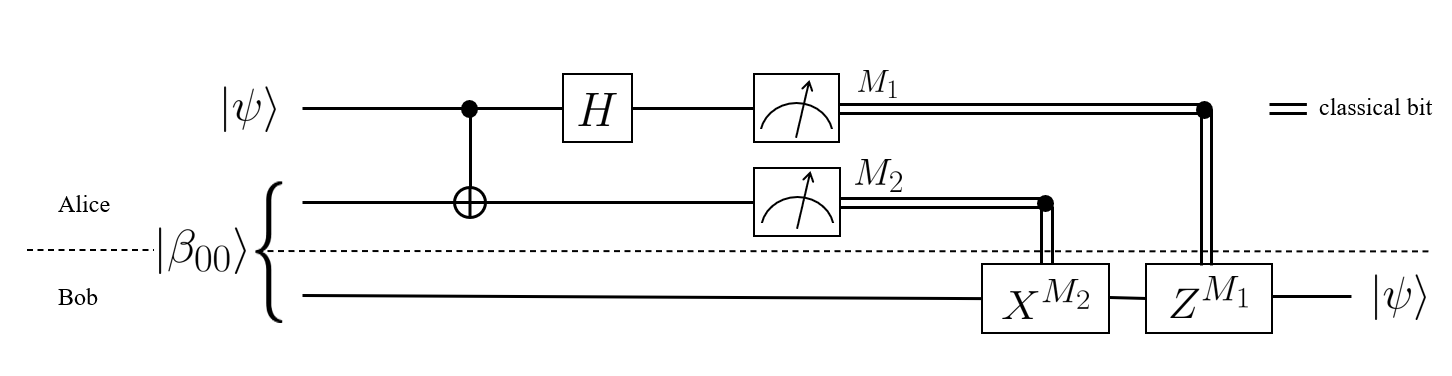}
\caption{Quantum circuit for teleporting a qubit}
\label{qu_teleport}
\end{figure}

Figure \ref{qu_teleport} shows the quantum circuit that implements the teleportation of a qubit. The state to be teleported is $\ket{\psi} = \alpha \ket{0} + \beta \ket{1}$, where $\alpha$ and $\beta$ are unknown amplitudes. The state input into the circuit is $\ket{\psi_0} = \ket{\psi} \ket{\beta_{00}}$. The bottom line represents Bob's system, while the two top lines are Alice's system. The single lines denote qubits, the meters represent measurement operations, and the double lines coming out of them carry represent classical bits.

There are two interesting features of teleportation. Quantum teleportation does not enable faster than light communication, because to complete the teleportation Alice must transmit her measurement result to Bob over a classical communication channel. Second, teleportation doesn't violate the no-cloning theorem, creating a copy of the quantum state being teleported. This violation is only illusory since after the teleportation process only the target qubit is left in the state $\ket{\psi}$, and the original data qubit ends up in one of the computational basis states $\ket{0}$ or $\ket{1}$, depending upon the measurement result on the first qubit.

Teleportation depends on the ability to create entangled Bell pairs over some distance. Naturally, many of the experimental groups involved in teleportation have also pushed the boundaries of what is possible to create larger, longer-distance, higher-fidelity, or longer-lived entangled states. In 1998, Boschi et al. verified the initial boundaries of teleportation \cite{PhysRevLett.80.1121}. The distance was increased in August 2004 to 600 meters, using optical fiber \cite{Ursin2004}. Subsequently, the record distance for quantum teleportation has been gradually increased to 16 kilometers (9.9 mi) \cite{Wei2010}, then to 97 km (60 mi), and after to 143 km (89 mi), set in open-air experiments in the Canary Islands, done between the two astronomical observatories of the Instituto de Astrof\'{i}sica de Canarias \cite{Ma2012}. Takesue et al. reached the distance of 102 km (63 mi) over optical fiber in 2015 using superconducting nanowire detectors \cite{Takesue:15}. The group of Jian-Wei Pan \cite{pan2017} reported the distance of 1,400 km (870 mi) by using satellite for space-based quantum teleportation, point out need to distribute Bell pairs across distances motivated in part by teleportation.

\subsection{Entanglement swapping} \label{swapp}
A simple and illustrative example of teleportation is the entanglement swapping, which will be seen to be very important for networking. The term "entanglement swapping" was introduced by Zukowski et al. \cite{Zukowski1993}, originally in the context of photonic Bell pairs created via PDC and coupled using beamsplitters. However, the entanglement swapping can be used to distribute Bell pairs across long distance by teleporting the state of one member of a Bell pair over progressively longer distances until the pair stretches from end to end. 

\begin{figure}[h!t]
\centering
\includegraphics[width=10cm, height=10cm]{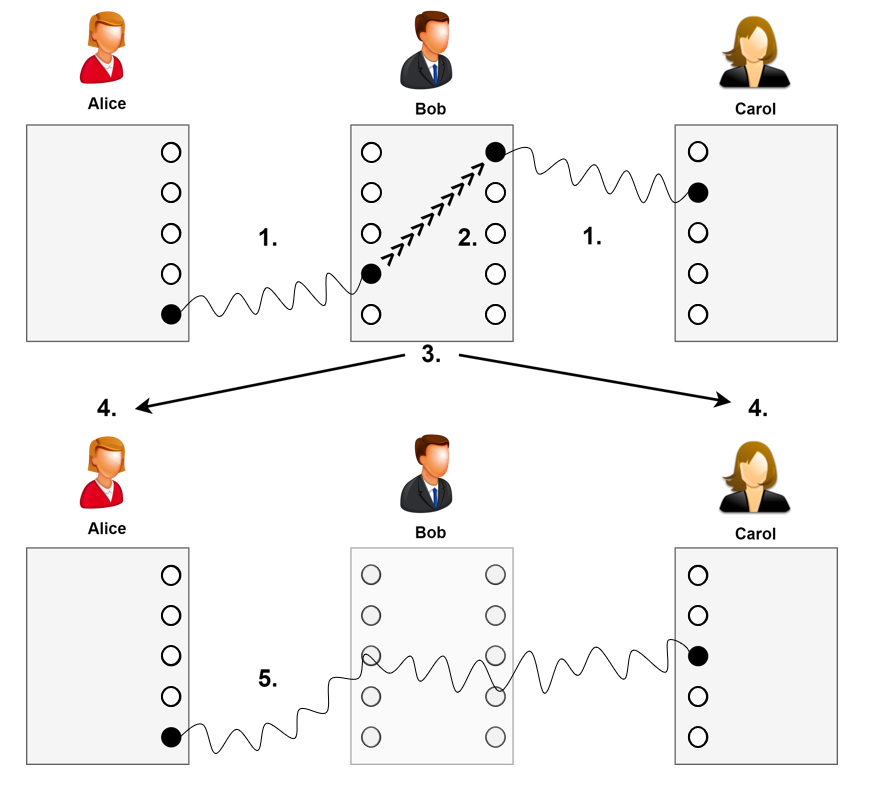}
\caption{Basic Entanglement Swapping}
\label{ent_swapp}
\end{figure}

Suppose Alice has a particle which is entangled with a particle owned by Bob, and Bob teleports it to Carol, then afterward, Alice's particle is entangled with Carol's. Figure \ref{ent_swapp} illustrates the process. Bob holds one end of each of two Bell pairs, one coupled to a qubit with Alice, the other to a qubit with Carol, which we will call $\ket{\psi^-}^{(AB)}$ and $\ket{\psi^-}^{(BC)}$ respectively. Bob decides to lengthen the pair on the left using the pair on the right. The results of this operation must be communicated to Carol, allowing Carol to recreate the state of  resulting in a new Bell pair.

In theory, Alice never needs to be told that the operation has occurred. Although Carol must apply corrective operations to complete the reconstruction, Alice is entirely passive, merely storing its half of the Bell pair in a buffer memory. However, Alice is very likely waiting on the completion of the swapping operation in order to perform some other action; at the very least, an application at node Alice is waiting to use the end-to-end Bell pair. 

\subsection{Purification and error correction} \label{purif_correct}
According to Van Meter, purification is the process of improving our knowledge about the state, by testing propositions about it \cite{VanMeter2014quantum}. This improvement is reflected as an increase in the fidelity in the density matrix that represents our knowledge about the state. We can describe a purification protocol in terms of: 
\begin{enumerate}
 \item the number and type of input states;
 \item the test procedure for certain propositions; %, consisting of
 \item the scheduling algorithm used to select states for participation in purification.
\end{enumerate}
In general, the required input states are two imperfect Bell pairs, with the goal being to produce one output Bell pair of higher fidelity. 

\begin{figure}[h!t]
\centering
\includegraphics[]{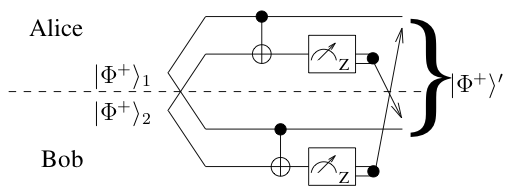}
\caption{Circuit for basic purification according \cite{VanMeter2014quantum}. }
\label{fig_purific}
\end{figure}

To illustrate, we can consider the circuit for basic purification of Figure \ref{fig_purific}, used by Van Meter \cite{VanMeter2014quantum}. Classical messages exchanged between Alice and Bob are indicated by the arrows.  

When the Bell pairs, the gates, and measurements are all perfect, the two $CNOT$ gates cancel, as illustrated in Equation (\ref{Purification}) and we still have two unentangled $\ket{\Phi^+}$ pairs. Alice and Bob each have a 50\% chance of finding 0 and 50\% chance of finding 1 when the second pair is measured, and when they exchange their measurement results they will always find the same value.

\begin{eqnarray} \label{Purification}
\ket{\Phi^+}_1 \ket{\Phi^+}_2 & = & (\ket{00} + \ket{11})  (\ket{00} + \ket{11})
\end{eqnarray}
\begin{eqnarray*}
& = & \ket{00} \ket{00} + \ket{00} \ket{11} + \ket{11} \ket{00} + \ket{11} \ket{11} \\
&   & \underrightarrow{CNOT_A} \ket{00} \ket{00} + \ket{00} \ket{11} + \ket{11} \ket{10} + \ket{11} \ket{01} \\
&   & \underrightarrow{CNOT_B} \ket{00} \ket{00} + \ket{00} \ket{11} + \ket{11} \ket{00} + \ket{11} \ket{11} \\
&  = & \ket{\Phi^+}_1 \ket{\Phi^+}_2  
\end{eqnarray*} 

When a Bell pair suffer a bit flip error, $\rho = P \ket{\Phi^+} \bra{\Phi^+} + (1 - P ) \ket{\Psi^+} \bra{\Psi^+}$, we would find that is indeed a $\ket{\Phi^+}$ pair, with probability $P$, if we could test pair 1 directly, and we would find that it is a $\ket{\Psi^+}$ pair, with probability $1 - P$. In the case when both pairs are in fact in $\ket{\Phi^+}$, the circuit operates as presented in Equation (\ref{Purification}) and the probability it happens is $P^2$.

If either of the Bell pairs has a bit flip error, Alice and Bob will find different values when they measure their qubits and we have to discard pair $1$, even though it might be good because we cannot tell if the error was in pair $1$ or pair $2$. If both Bell pairs have an error, Alice and Bob will find the same value. With probability $(1 - P)^2$ , the error in pair $1$ goes undetected due to the error in pair $2$.

The probability of successful operation is $P^2 + (1 - P )^2$, including the false positive case engendered by two errors, while the probability of operation failure is $2P (1 - P )$. The resulting fidelity $(F_{ap})$ when operation succeed is

\begin{equation}
F_{ap} = \frac{P^2}{P^2 + (1-P)^2}
%    \ket{\psi} = \cos {\frac{\theta}{2}}\ket{0} + e^{i\varphi} \sin {{\frac{\theta}{2}} \ket{1}}.
\end{equation}
with the following final state
\begin{equation}
\rho_{ap} = F_{ap} \ket{\Phi^+} \bra{\Phi^+} + (1 - F_{ap}) \ket{\Psi^+} \bra{\Psi^+}
\end{equation}

\begin{figure}[h!t]
\centering
\includegraphics[]{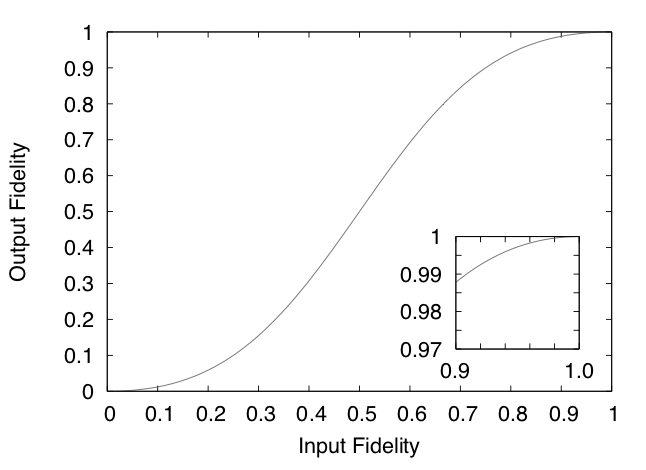}
\caption{Output fidelity as a function of input fidelity according \cite{VanMeter2014quantum}. }
\label{fig_fidelity}
\end{figure}

To check the quality of resulting fidelity is common to analyze the output fidelity as a function of input fidelity. Van Meter \cite{VanMeter2014quantum} performed this analysis for basic purification of two identical Bell pairs with bit flip errors only, and perfect purification operations, as illustrated in Figure \ref{fig_fidelity}. $F_{ap} > F$ when $F > 0.5$ and for input fidelity greater than approximately 0.8, the improvement in fidelity is very good.

Various techniques for managing errors have been developed, some based on classical error correction and erasure correction techniques, others on uniquely quantum approaches \cite{Devitt_2013} \cite{Terhal_2015}. Purification, in which two or more multiqubit states are manipulated to form one higher-fidelity state, uses few quantum memory resources and simple quantum operations, but operates only on well-understood states such as Bell states rather than arbitrary application data. 

\subsection{Quantum repeaters} \label{q_repeat}
A fundamental component of a quantum network is the quantum repeater. It allows the transportation of qubits over long distances, which is hindered by signal loss and decoherence inherent to most transport mediums such as optical fiber. Since absorption losses and depolarization error scale exponentially as distance increases, one cannot hope to cover any long distance between A and B in one leap \cite{decayloss2017}. Repeaters appear in-between end nodes. Loss in telecommunications fiber is typically around 0.2 dB/km; high-quality fibers with loss of 0.17dB/km are available, and in laboratories, loss is as low as 0.12 dB/km. At 0.17 dB/km, the attenuation length is about 25 km. This value is commonly used in quantum repeater simulations, although NTT (Nippon Telegraph and Telephone) accomplished the remarkable feat of distributing time-bin entangled photons through 300 km of fiber \cite{Inagaki_2013}. 

In classical communication, amplifiers can be used to boost the signal during transmission, but in a quantum network amplifiers cannot be used since qubits cannot be copied - known as the no-cloning theorem. By necessity, a quantum repeater works in a fundamentally different way than a classical repeater. Quantum repeaters allow entanglement and can be established at distant nodes without physically sending an entangled qubit the entire distance \cite{Bouwmeester_1997}. A quantum repeater protocol executes three operations to create the long-range Bell state that can be used for quantum communication tasks such as QKD or teleportation. These operations are:

\begin{itemize}
    \item Entanglement distribution: the process for creating entangled links between network nodes, as presented in Section \ref{telep}.
    \item Entanglement purification: the process where we create a more highly entangled state from a number of lower quality ones, as described in Section \ref{purif_correct}.
    \item Entanglement swapping: the process in which a Bell-state measurement is performed within a node on two qubits which are halves of separate Bell states, as illustrated in Section \ref{swapp}. The Bell measurement allows us to generate a longer entangled link connecting adjacent repeater nodes.
\end{itemize}

The first operation is needed only between shorter-range adjacent nodes and thus the success probability for generating the entangled link depends on the distance of the adjacent nodes, rather than the total communication distance. The second operation has the objective to not permit information present in the state has been lost. The third operation is the mechanism to extend the range of the entanglement.

While the quantum repeater protocol for generating long-range entanglement may seem quite straightforward in nature, its behavior is quite complex due to the various probabilistic elements inherent in the scheme \cite{sangouard2009quantum}. Different generations of quantum repeaters have already been developed \cite{Inagaki_2013} \cite{7010905} and progress has been significant in recent years both from an engineering perspective but also with new approaches \cite{muschik2019}. As the performance of these systems continues to improve they will also be able to take advantage of the developments in QKD, and secure quantum communication in general, in terms of their integration in standard fiber-optic networks. 

Early architectures of repeaters, in essence, used teleportation to extend entanglement, and purification to detect errors introduced in the process \cite{munro2013quantum}. The process of entanglement swapping uses teleportation to splice two Bell pairs spanning adjacent short distances into one pair over the corresponding longer distance, as presented in Figure \ref{ent_swapp}. Entanglement swapping is independent of the distances between A and B, and between B and C. Only local quantum operations are required, supported by classical communication. Purification is used to compensate for the errors introduced, as described in Section \ref{purif_correct}. Local quantum operations are performed at both nodes on two Bell pairs, then one of the Bell pairs is measured. The measurement results are exchanged and compared. If they agree, the pair's fidelity has improved, and it is kept for reuse. If the measurement results disagree, the pair is discarded.

\begin{figure}[h!t]
\centering
\includegraphics[width=8cm, height=8cm]{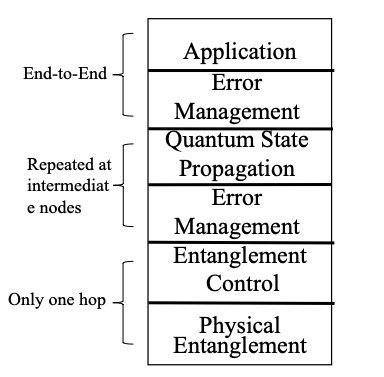}
\caption{Basic protocol stack architecture for networks of quantum repeaters.}
\label{fig_prot_stack}
\end{figure}

From these concepts to execute a distributed algorithm that creates entangled quantum states between nodes that are far apart, different researchers \cite{Jiang_2009} \cite{METER_2011} presented simple protocol stacks for networks of quantum repeaters that considers all the necessary classical messages and which can be easily adapted for different approaches, as illustrated in Figure \ref{fig_prot_stack}.

The physical entanglement layer represents the physical interaction that creates Bell pairs between two different stations. Many technologies for this layer are under development \cite{munro2013quantum}. The second layer, Entanglement Control, is responsible for managing the single-hop physical entanglement process, selecting qubits to attempt entanglement at each end of the link, and utilizing classical messages to report the results. The third layer of the protocol, Error Management, is responsible for choosing two Bell pairs, and electing one pair to have its fidelity boosted and the other to be sacrificed, assuring that both stations make the same decisions. The fourth layer, Quantum State Propagation, is responsible for administering the Bell pairs, especially for networks with shared resources. Important decisions, such as whether to purify or swap first or when to swap, need to be carefully taken. The Application layer will determine if end-to-end entanglement is required, or if our quantum states can be measured on a pay-as-you-go basis. Currently, the most important existing application is QKD (Quantum Key Distribution). However, the application may be a sensor network, or a numeric computation or decision algorithm based on shared state, as we will see in Section \ref{q_app}.

It is important to mention that QKD applications can be performed not only creating an end-to-end quantum channel but also using trusted repeaters. Consider two end nodes Alice(A) and Bob(B), and a trusted repeater R in the middle. A and R now perform quantum key distribution to generate a key $K_{AR}$. Similarly, R and B run quantum key distribution to generate a key $K_{RB}$. A and B can now obtain a key $K_{AB}$ between themselves as follows: A sends $K_{AB}$ to R encrypted with the key $K_{AR}$. R decrypts to obtain $K_{AB}$. R then re-encrypts $K_{AB}$ using the $K_{RB}$ and sends it to B. B decrypts to obtain $K_{AB}$. A and B now share the key $K_{AB}$. The key is secure from an outside eavesdropper, but clearly the repeater R also knows $K_{AB}$. This means that any subsequent communication between A and B does not provide end to end security, but is only secure as long as A and B trust the repeater R. 

A true quantum repeater allows the end to end generation of quantum entanglement, and thus - by using quantum teleportation - the end to end transmission of qubits. In quantum key distribution protocols, one can test for such entanglement. This means that when making encryption keys, the sender and receiver are secure even if they do not trust the quantum repeater. Any other application of a quantum Internet also requires the end to end transmission of qubits, and thus a quantum repeater. 

To make maximum use of communication infrastructure, we also require optical switches capable of delivering qubits to the intended quantum processor. These switches need to preserve quantum coherence, which make them more challenging to realize than standard optical switches. Current commercial switches have various problems that make them unsuitable for rerouting entangled photons. Those that are made of micro-electromechanical components keep entangled states intact but operate too slowly. Other optoelectronic switches either add too much noise so that single photons are difficult to detect, or they completely destroy the quantum information. The utilization of a quantum switch provides significant advantages for a number of problems, ranging from quantum computation and quantum information processing, through non-local games to quantum communication \cite{Caleffi_2020}. 

\subsection{Quantum Key Distribution} \label{q_app}
In literature, there are several types of quantum applications. They are often classified into two categories, distributed agreement protocols, and distributed computation, although the underlying theory is essentially the same \cite{VanMeter2014quantum}. However, in this section, our focus will be on Quantum Key Distribution (QKD), the most practical, commercially attractive use of quantum networks in the near term. We will present how the protocol works and the network requirements to support application. 

The main objective in QKD is the use of quantum mechanics to detect the presence or absence of an eavesdropper. QKD systems generate shared, secret random numbers between two distant parties. Shared random numbers, if provably secret, can be used as cryptographic keys, allowing secure communication across physically insecure networks such as the Internet. An important and unique property of QKD is the ability of the two communicating users to detect the presence of any third party trying to gain knowledge of the key. This results from a fundamental property of quantum mechanics: the process of measuring a quantum system, in general, disturbs the system. A third party trying to eavesdrop on the key must in some way measure it, thus introducing detectable anomalies. By using quantum superpositions or quantum entanglement and transmitting information in quantum states, a communication system can be implemented that detects eavesdropping. If the level of eavesdropping lies below a certain threshold, a key can be produced that is guaranteed to be secure (i.e., the eavesdropper has no information about it), otherwise no secure key is possible and key generation is aborted. 

Artur Ekert proposed, in 1991, a QKD protocol currently called E91 \cite{E91prot}, using entangled pairs of photons. The entangled states are perfectly correlated in the sense that if Alice and Bob both measure whether their particles have vertical or horizontal polarizations, they always get the same answer with 100\% probability. The same is true if they both measure any other pair of complementary (orthogonal) polarizations. This necessitates that the two distant parties have exact directionality synchronization. However, the particular results are completely random; it is impossible for Alice to predict if she (and thus Bob) will get vertical polarization or horizontal polarization. Any attempt at eavesdropping by Eve destroys these correlations in a way that Alice and Bob can detect. The correct operation of E91 does require a functioning quantum network capable of generating the required Bell pairs. The fidelity is also important, because, with low fidelity, the eavesdropper detection becomes more difficult and consumes a larger fraction of the end-to-end Bell pairs.

QKD can be incorporated into a production classical network in different ways \cite{Pirker_2019}. A simple arrangement is to securely connect two networks in two far locations, that can belong to the same organization or not. One approach is to use a virtual private network (VPN) to connect the two locations \cite{VanMeter2014quantum}. Implementations of QKD are well beyond the experimental phase \cite{dodson2009updating}. A few commercial products are available, and metropolitan-area testbed networks exist in Boston, Vienna, Geneva, Barcelona, Durban, Tokyo, several sites in China and elsewhere throughout the world. In fact, the BB84 \cite{bennett1984QKD} technique deployed in most links in these networks does not use entangled quantum states, although another approach, developed by Artur Ekert, does \cite{PhysRevLett.67.661}. QKD has also been integrated into custom encryption suites and the Internet standard IPsec suite and has been proposed for use with the TLS protocol commonly used on the World Wide Web \cite{mink2010quantum}.

QKD can also be viewed as a form of sensor network: the goal of the underlying quantum operation is the same, physical detection of eavesdropping on the quantum channel \cite{VanMeter2014quantum}. Distributed entanglement is an extremely sensitive physical state, and can be used as a physical probe for other applications, like improving the resolution of optical telescopes using interferometry \cite{PhysRevLett.109.070503} and comparing the relative time of two clocks separated by a distance \cite{komar2014}. Both of these applications are far from practical given both the current state of the technology and the very demanding nature of the existing proposals, but they serve as important signposts on the road to the merger of quantum information and real-world sensors and actuators.

\subsection{Quantum networks and Quantum Internet} \label{q_internet}
We presented in the previous sections that quantum networks form an important element of quantum computing and quantum communication systems. Quantum networks facilitate the transmission of information in the form of qubits between physically separated end nodes. Quantum networks, like classical networks, will involve nodes and links and a layered communication architecture with individual protocol modules communicating vertically up and down a protocol stack and horizontally with peers. There are, however, fundamental differences that make the merger of classical and quantum networking concepts less than straightforward \cite{di2012optimal}. 

In principle, we can consider two main approaches to construct quantum networks \cite{sigcom2019}. On the first alternative, quantum networks could simply forward quantum information directly, which however needs to be protected against noise and decoherence using quantum error correcting codes, and repeatedly refreshed at intermediate stations where error correction is performed. On the second one, quantum networks may use entanglement, a property, as presented in Section \ref{multi_qbits}, that is only accessible in quantum physics. Constructing quantum networks by using entanglement has one significant advantage compared to the first approach: the entanglement topology of a network, which determines in that case also the boundaries and ultimately the structure of a network, is completely independent of the underlying physical channel configuration.

As shown in Section \ref{q_repeat}, a crucial element to establishing long-distance entanglement are quantum repeaters, and multiple proposals for repeater-based networks have been put forward \cite{METER_2011} \cite{rubino2017experimental}. Although most schemes are based on bipartite entanglement, where Bell pairs are generated between nodes of the network, future quantum networks shall not be limited to the generation of Bell-pairs only, because many interesting applications require multipartite entangled quantum states \cite{Pirker_2019}.

The basic structure of a quantum network and more generally a quantum Internet is analogous to a classical network \cite{quantumInternet2018}. Besides quantum repeaters, we have end nodes, quantum channels, and quantum switches. Applications run in end nodes. These end nodes can be quantum processors containing at least one qubit. Most applications of a quantum Internet require only very modest quantum processors. For most quantum Internet protocols, such as QKD, it is sufficient if these processors are capable of preparing and measuring only a single qubit at a time.  On the other hand, some applications of a quantum Internet require quantum processors of several qubits as well as a quantum memory at the end nodes. Quantum network nodes can exchange classical control information over standard classical communication channels. This may be by means of a direct physical connection or via, for example, the Internet.

Layering is a natural means of dividing functionality, and the associated modularity allows us to replace individual functions more or less independently. In this context, it is essential to develop methods that allow quantum protocols to connect to the underlying hardware implementation transparently and to make fast and reactive decisions for generating entanglement in the network in order to mitigate limited qubit lifetimes. However, only preliminary functional allocation of a quantum network stack has been proposed, and just first versions of physical and link layer protocols have been developed \cite{sigcom2019} \cite{Pirker_2019} \cite{VanMeter2014quantum}.

\begin{figure}[h!t]
\centering
\includegraphics[width=15cm, height=9cm]{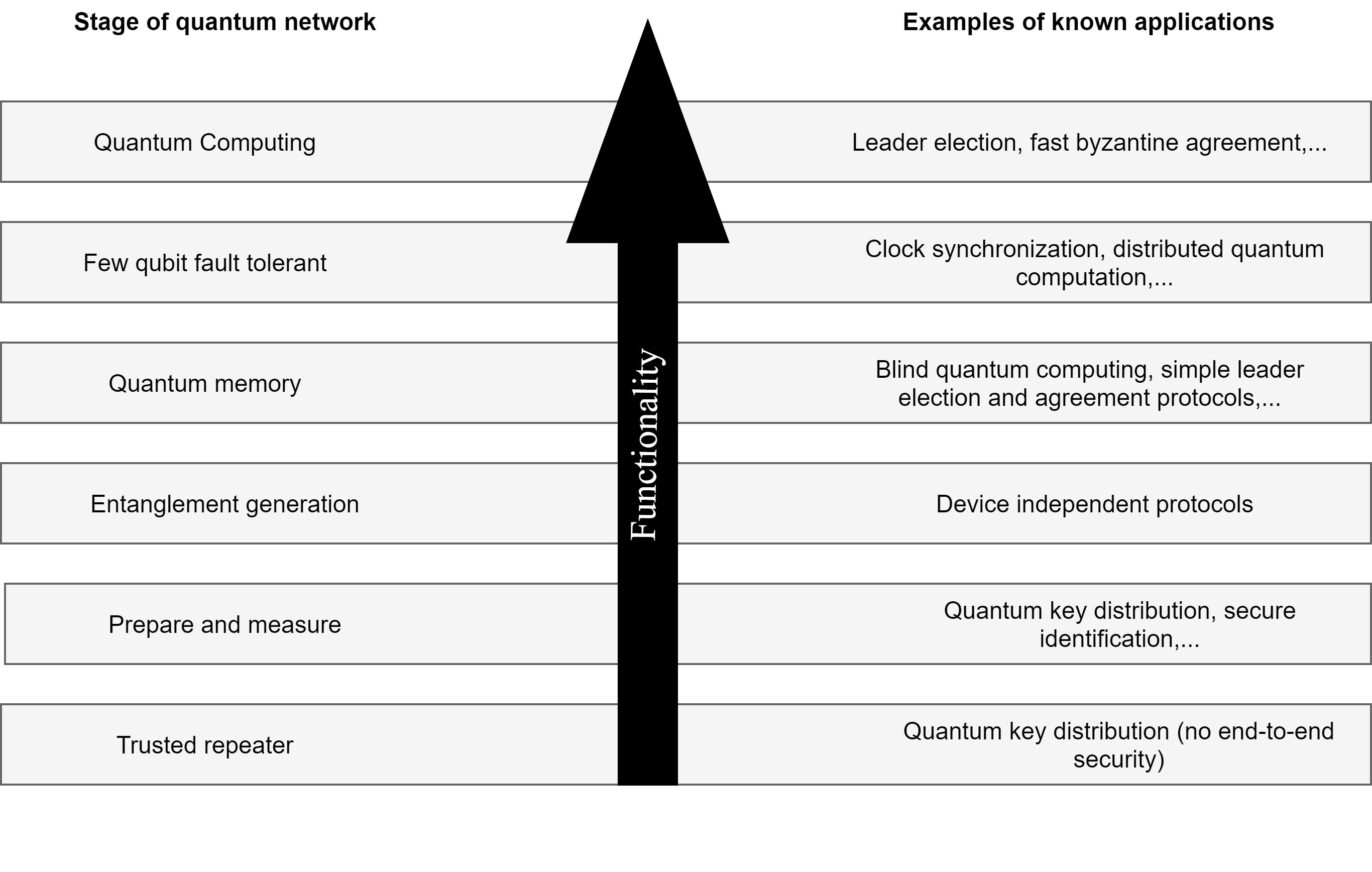}
\caption{Stages of Quantum Internet according Stephanie Wehner et al. \cite{Wehner2018vision}.}
\label{Internet_stages}
\end{figure}

Stephanie Wehner et. al \cite{Wehner2018vision} propose stages of development toward a full-blown quantum Internet. They suggested stages that are functionality driven: Central to their definition is not the difficulty of experimentally achieving them but rather the essential question of what level of complexity is needed to actually enable useful applications. Each stage is interesting in its own right and distinguished by a specific quantum functionality that is sufficient to support a certain class of protocols, as illustrated in Figure \ref{Internet_stages}. \\

Each stage is characterized by an increase in functionality at the expense of greater technological difficulty. A specific implementation of a quantum Internet may, like for a classical network, be optimized for distance, functionality, or
both. The objective of a network is to provide any end nodes (connected to the network) with the means to exchange data, making three end nodes the smallest instance of a true network. 

We will briefly describe only the first two stages because they are ones that have been realized in practice in some way. A trusted repeater network,  first stage, has at least two end nodes and a sequence of short distance links that connect nearby intermediary repeater nodes. Each pair of adjacent nodes uses QKD to exchange encryption keys. The main characteristic of trusted repeater networks is that they do not allow the end-to-end transmission of qubits. Currently, outside the laboratory, only trusted repeater networks have been realized in metropolitan areas \cite{Wang_2014}. 

The second stage, prepare and measure networks, enables end-to-end QKD without the need to trust intermediary repeater nodes and already allows other protocols. It allows any node to prepare a one-qubit state and transmit the resulting state to any other node, which then measures it. This stage is the first to offer end-to-end quantum functionality and demands the use of quantum repeaters to bridge long distances via intermediate qubit storage or error correction, as well as routers to forward the quantum state to the desired node. Several recent experiments have demonstrated elements belonging to this stage \cite{Pirker_2019}

In 2011, Van Meter et. al \cite{METER_2011} introduced the concept of a Quantum Recursive Network Architecture (QRNA), developed from the emerging classical concept of recursive networks, extending recursive mechanisms from a focus on data forwarding to a more general distributed computing request framework. Recursion abstracts independent transit networks, as single relay nodes, unifies software layering and virtualizes the addresses of resources to improve information hiding and resource management. The architecture is useful for building arbitrary distributed states, including fundamental distributed states such as Bell pairs and GHZ, W and cluster states.

Routing is another interesting aspect of quantum networks \cite{routing2019}. Routing messages to the right destination in a network has seen enormous attention in the classical literature, and the term routing is also used for a number of different concepts in quantum networks. It is highly likely that concepts from both domains will form an important ingredient in designing quantum networks \cite{chakraborty2019_routing}.

Applying the appropriate simplifications, routing in a quantum network can be understood as routing on virtual quantum links (VQLs). However, such VQLs are rather unusual from a classical perspective in that they can be used only once, and require one qubit of quantum memory at each endpoint to be maintained \cite{Don2019}. While the requirements of a VQL for one qubit of memory at each node appears benign from a classical perspective, it is rather significant in a quantum network. First of all, due to current technological limitations, each network node can store only very few qubits. What's more, such quantum storage is typically rather noisy, meaning that each qubit has a limited lifetime. The latter can, in theory, be overcome by performing error-correction at each network node at the expense of using additional qubits. In resume, the VQLs can be assigned deliberately and dynamically, carry a certain cost, can only be used once, and which may expire after a given time.

As a result, communication in a quantum network can, in principle, be understood entirely as transformations performed on the graph of VQLs: Given existing VQLs, to send a qubit from two nodes A and B we consume VQLs to create a new entangled link - a new VQL - directly between A and B, followed by teleportation of the qubit over said VQL \cite{schoute2016shortcuts}.

\section{Current Scenario and Research Challenges} \label{current_scen}
The  Quantum  Internet  is  envisioned  as  the  final stage  of  the  quantum  revolution,  opening  fundamentally  new communications  and  computing  capabilities,  including  distributed   quantum   computing. This section presents the current status of the quantum Internet, along with challenges, and research opportunities in this emerging area. The focus is on laying the groundwork to adapt Internet design principles to the development of quantum networks. We discuss the key research challenges and open problems related to the design of a quantum network, which harnesses quantum phenomena, such as entanglement and superposition, to share quantum states among remote quantum devices.

\subsection{Current scenario}
The quantum Internet describes a collection of distributed quantum nodes, separated by a range of distances over which one desires to perform some quantum communication protocol that can support, for example, distributed quantum computation or distributed sensing \cite{Pirandola2016}. There are currently numerous quantum communication and cryptographic protocols identified, including security distribution for encryption \cite{mink2010quantum, Kumar_2019}, quantum-certified random number generation in the form of random number beacons and personal devices, secret-sharing \cite{Hillery_1999, Williams_2019}, quantum fingerprinting \cite{Guan_2016} and other multi-party computation protocols, such as secure quantum voting, byzantine agreements, and multi-party private auctions \cite{Broadbent_2015}. 

As we mentioned in Section \ref{q_internet}, the current experimental status of long-distance quantum networks is at the lowest stage with several commercial systems for QKD on the market. The first extended trusted repeater networks have already been implemented over metropolitan distances \cite{Wang:14}, and a long-distance implementation has recently been completed \cite{Kumar_2019}. The hardware required for the lowest stage has been described in detail in \cite{Diamanti2016}. Realizing the next stages with end-to-end quantum functionality over long distances demands, basically, the use of quantum repeaters to bridge long distances via intermediate qubit storage or error correction, as well as routers to forward the quantum state to the desired node. Several recent experiments have demonstrated elements belonging to this and higher stages at short distances, suggesting that higher-functionality networks are within reach \cite{Cacciapuoti_2020}. To put these advances into the right perspective, we briefly summarize the main requirements for the essential quantum Internet hardware.

Quantum links between the repeater stations and the end nodes are established via photonic channels. Two types of photonic channels can be considered: free-space channels, potentially via satellites \cite{Yin2017}, and fiber-based channels. Each has its own advantages and disadvantages, and a future quantum Internet may use a combination of them, similar to the current classical Internet. Hybrid architectures will probably be used for connections faring both the use of cryo-cables (expensive and necessarily limited in length) and of optical fibers or free space photonic links. We require these channels to exhibit minimal photon loss and decoherence. 

The end nodes need to meet the following requirements for the quantum Internet to reach its full potential:
\begin{itemize}
    \item Robust storage of quantum states during the time needed to establish entanglement between end nodes. This robustness must persist under quantum operations performed on the end node.
    \item High-fidelity processing of quantum information within the node. For more advanced tasks, multiple qubits will be required, making the end nodes similar to small-scale quantum computers.
    \item Compatibility with photonic communication hardware: efficient interface to light at the relevant wavelength.
\end{itemize}

Several experimental platforms are currently being pursued for the end nodes. Each of these combines well-controlled matter-based qubits with a quantum optical interface via internal electronic transitions. 

For quantum repeaters, the requirements are less strict than for the end nodes. Depending on the architecture of the repeaters, the storage of quantum states may only be required for the time needed to establish entanglement between the nearest active nodes, substantially different from the storage time required for the end nodes. Also, the qubit processing capabilities required are limited, and therefore systems different from the end nodes can be considered. 

As mentioned in Section \ref{q_repeat}, the quantum switch is another essential component for a quantum Internet. Multiple physical implementations of the quantum switch have been proposed and experimentally realized with photons, with the control qubit represented by polarization or orbital angular momentum degrees of freedoms. Preliminary results of the quantum switches to face with the noise degradation introduced by the entanglement distribution are very good. However, a substantial amount of conceptual and experimental work has to be developed in order to tackle the challenges and open problems associated with the utilization of the quantum switch in the Quantum Internet  \cite{rubino2017experimental} \cite{awschalom2019development}.

At the present time, however, quantum networking in the real world consists of three research programs and commercialization efforts: the first one is Quantum Key Distribution (QKD) that adds unbreakable coding of key distribution to public-key encryption, as we presented in Section \ref{q_app}. 

The second one is cloud/network access to quantum computers. It is core to the business strategies of leading quantum computer companies. Cloud-based quantum computing is the invocation of quantum emulators, simulators or processors through the cloud. Increasingly, cloud services are being looked on as a method for providing access to quantum processing. IBM already had connected a small quantum computer to the cloud and it allows users to execute simple programs on the cloud \cite{Ibmcloud}. Many people from academic researchers and professors to schoolkids have built programs that run many different quantum algorithms using the program tools. Some consumers hope to use fast computing to model financial markets or to build more advanced AI systems \cite{chen2018experimental}. 

The last one is quantum sensor networks, which could exploit the correlations across an array of sensors, linking them to each other with quantum mechanical means \cite{Degen_2017}. Quantum sensors exploit superposition, entanglement, squeezing, and backaction evasion to make measurements with a precision better than the Standard Quantum Limit (SQL), with the ultimate goal of reaching the Heisenberg Limit. Sensor networks improve the sensitivity and scalability of the resulting entangled system simultaneously allowing it to benefit from the long-distance baseline between the sensors. Some promising options are quantum networks of atomic clocks, phase-sensitive quantum networks, quantum networks of magnetometers \cite{proctor2017networked}.

For the past 15 years, major service providers and research institutions worldwide have run quantum network trials. We are now entering a period in which permanent quantum networks are being built. These are designed initially to support quantum encryption services, but will soon also provide the infrastructure for quantum computing. As quantum networks are deployed, they will eventually create opportunities at the service level, but more immediately at the components and modules level. This is because quantum networks will require a slew of new optical networking technologies to make them function effectively. The market for quantum networking is projected to reach \$5.5 billion by 2025, according to a new report from Inside Quantum Technology (IQT) \cite{IQT2020}. 

\subsection{Research opportunities}
In this subsection, we discuss the key research challenges and open problems related to the design of a quantum network, which harnesses quantum phenomena with no-counterpart in the classical reality, such as entanglement and teleportation, to share quantum states among remote quantum devices.

Exploring how to build the quantum Internet, a vast network of quantum computers and other quantum devices, will catalyze new technologies that accelerate today's Internet, improve the security of our communications, and allow dramatic advances in computing \cite{Caleffi_2020}. The difficulty of every item in the design of a network grows as the scale of the network increases. This is true for both, classical and quantum networks. The main challenges in scaling networks to Internet-scale and beyond are: heterogeneity, especially of deployed technologies and local conditions; sheer scale, affecting routing and naming in particular; dealing with out of date information about current network conditions (e.g. routing or congestion) and the success or failure of requested operations; meeting the needs of participating organizations, such as privacy, desired traffic transit policies and autonomous management; and dealing with misbehaving nodes on the network, whether the misbehavior is deliberate or accidental \cite{Cacciapuoti_2020}.

As we presented in Section \ref{q_bits} qubits are very fragile: any interaction of a qubit with the environment causes decoherence, i.e., a loss of information from the qubit to the environment as time passes, and isolation is hard to achieve in practice given the current state-of-the-art of quantum technologies. Furthermore, perfect isolation is not desirable, since computation and communication require interaction with the qubits, e.g., for reading/writing operations. Although a gradual decrease of the decoherence times is expected with the progress of the quantum technologies, the design of a quantum network must carefully account for the constraints imposed by quantum decoherence.

Decoherence is not the only source of errors. Errors practically arise with any operation on a quantum state due to imperfections and random fluctuations. Here, a fundamental figure of merit is the quantum fidelity, presented in Section \ref{Interf_decoh_fidelity}. From a communication engineering perspective, the joint modeling of errors induced by the quantum operations, together with those induced by entanglement generation/distribution, is still an open problem.

Furthermore, the no-cloning theorem prevents the adoption in quantum networks of classical error recovery techniques, depending on information cloning, to preserve quantum information against decoherence and imperfect operations. Recently, many quantum error correction techniques have been proposed as in \cite{8218756}. However, further research is needed. In fact, quantum error correction techniques must handle not only bit-flip errors, but also phase-flip errors, as well as simultaneous bit- and phase-flip errors. This differs from classical networks that only have to consider the bit-flip error. 

One fundamental difference with respect to classical networks, where broadcast is widely exploited for implementing several link layer and network layer functionalities, such as medium access control and route discovery is the impossibility of transmitting quantum information to more than a single destination due to no-broadcasting theorem \cite{Barnum_2007}, a corollary of the no-cloning theorem. As a consequence, the link layer must be carefully re-thought and re-designed, and effective multiplexing techniques for quantum networks should be designed to allow multiple quantum devices to be connected to a single quantum channel (e.g. a fiber). Access to the medium could be based for example on photon-frequency-division for the entanglement distribution.

Entanglement distribution determines the connectivity of a quantum network in term of capability to perform teleporting among quantum devices. Hence, novel quantum routing metrics are needed to ensure effective entanglement-aware path selection. Furthermore, the teleportation process destroys entanglement as a consequence of the BSM at the source. Hence, if additional qubits need to be teleported, new entangled pairs need to be created and distributed between the source and the destination. This constraint has no-counterpart in classical networks and it must be carefully accounted for in an effective design of the network layer \cite{Don2019, Kumar_2019}.

In relation to the deployment of a quantum Internet, there are several challenges for the near future. At first, quantum computers will be available in few, highly specialized, data centers capable of providing the challenging equipment needed for quantum computers. Companies and users will be able to access quantum computing power as a service via cloud. In this regard, the quantum cloud market is estimated nearly half of the whole 10 billion dollar quantum computing market by 2024 \cite{ibm2018}. IBM already allows researchers to design and execute quantum algorithms through classical cloud access to isolated 5-, 16- and 20-qubits quantum devices.  

To extend the range of fiber-based entanglement distribution beyond a few hundred kilometers, quantum repeaters are required. One important benchmark for quantum repeaters is the repeater-less bound, which imposes the fundamental limit of the direct quantum communication protocols. Recently, there have been significant advances in experimentally demonstrating key elements of a quantum repeaters in an integrated system. An important recent highlight is the experimental demonstration of memory-enhanced quantum communication surpassing repeater-less bound in a proof-of-concept laboratory setting \cite{Havard_MIT_qrepeater_2020} \cite{Pirandola_2017}. This paves the way towards the demonstration of a full quantum repeater, which in turn will enable scalable large-scale quantum networks.

An important challenge in extending the point-to-point entangled links into true networks is the problem of efficient storage of quantum states \cite{Hucul_2014}. Ideally, it is necessary to have a quantum memory that is capable of storing and releasing quantum states on the level of individual qubits and on-demand. Storage and on-demand retrieval have already been achieved \cite{Delteil_2017}, although efficiencies are still to be improved. 

Another challenge is that most of the above systems do not intrinsically couple to light in the telecom band. To fulfill the compatibility requirement with photonic communication hardware, wavelength conversion at the single-photon level can be used \cite{Zaske_2012}. While existing quantum interfaces between modules have seen dramatic improvements, most systems still have not reached the regime where connection between the modules can be utilized for reliable transfer of qubits within the timescale required for distributed quantum computation. The first realizations of a Quantum Internet probably will be small clusters of quantum processors within a data center. Architectures will have to take into account the high cost of data buses (economically and in terms of quantum fidelity) limiting both the size of the clusters and the use of connections for processing.

Seamless integration of the communication interfaces with the computational functions of the modules can also introduce some challenges. For example, in heralded entanglement generation protocols, the qubit-photon entanglement generation protocols can lead to decoherence of nearby qubits storing information. For these systems, novel integration approaches must be developed so that communication and local data processing can co-exist. For solid-state qubits (such as superconducting qubits) that use photons in the microwave range of the electromagnetic spectrum, communication over room-temperature channels becomes impractical. Given the fragile nature of quantum entanglement and the challenges posed by the sharing of quantum resources, a substantial amount of research is needed in the development both of novel networking protocols and of quantum and classical algorithms \cite{muschik2019}.

The ability to control, optimize, and recover from failures are critical in the design and operation of large-scale networks \cite{awschalom2019development}. Achieving these capabilities is a challenge that typically requires networks to carry network management and control information in addition to user data. There are two distinct methods to implement this: in-band control, if control and management information is carried on the same channel as user data, and out-of-band, if it is carried on a separate channel. Clearly, quantum networks need out-of-band control and signaling, since any attempt to read and process control information carried in the quantum channel will destroy its content.  

Modeling and performance analysis are important, both in the design phase to evaluate and compare the merits of a variety of quantum network protocols and architectures, as well as for real-time performance analysis and troubleshooting after the network is built \cite{vardoyan2019stochastic,vardoyan2019capacity}. Simulations will be needed to study network properties including quantum state and entanglement throughput, latency, scalability, reliability, and availability. One interesting alternative is NetSquid \cite{netsquid2020}, capable of simulating the decay of quantum information over time together with noisy operations and stochastic feedback loops. Since generic quantum systems consisting of even hundreds of qubits cannot be fully simulated on classical computers, ways to effectively employ reduced models are required. Some methods already exist that will clearly be useful, such as Monte Carlo simulations of systems whose operations only include Clifford gates. Some questions, for example, the extent to which various protocols are able to avoid bottlenecks, will be able to be addressed by purely classical simulations, using methods similar to those already developed by the classical networking community. Other questions, pertaining to the physical layer, will require simulation of the dynamics of optical channels and their interaction with the systems that comprise sending and receiving circuits. Such simulations will likely require the use of much more sophisticated methods such as matrix-product-state methods or tensor-network methods.

Quantum networks are complex, challenging engineered systems that require sophisticated solutions for their operations and control, with many of those solutions yet to be developed \cite{Caleffi_2020}. Indeed, many of the control plane technologies in use in modern classical networks are not suitable for the quantum data plane that cannot be subjected to O-E-O conversion, as discussed above. Quantum network management and operation will be particularly challenging due to the quantum nature embedded in the control plane and/or the data plane. The task is further complicated by the need for quantum networks to co-exist with conventional networks. In addition, monitoring of quantum networks requires measurements of complex conventional and quantum signals, along with inferences and analytics to distill knowledge and make control decisions \cite{ndoussefetter2019quantum}.

There are unique and extremely important security questions involved in the reliable, trustworthy operation and control of quantum resources to support quantum computation and quantum sensing efforts \cite{Cacciapuoti_2020}. Within the framework of coexistence infrastructures, security vulnerabilities of conventional networks carry over, and those of (newer) quantum components need to be explored and addressed. Furthermore, novel crossover vulnerabilities may potentially exist, wherein one modality may be exploited to compromise the other. Indeed, these aspects must be addressed from the start as an integral part of the design and analysis of quantum networks.

Another interesting research opportunity is the use of software defined network (SDN) technologies \cite{mckeown2009software} \cite{Aguado_2019}. In particular, the success of QKD networks radically depends on the degree to which they can be adopted in the existing infrastructure. The integration of  SDN, through the development of standard protocols and interfaces, has allowed new services and systems to be seamlessly integrated in telecommunications networks. The flexibility brought by SDN reduces drastically the effort of integrating new devices and technologies in the network and allows address the design of versatile quantum networks through the development of programmable quantum switches \cite{SDN_qsw2018}. 

Finally, quantum teleportation requires the integration of classical and quantum communication resources. Classical communication resources will be likely provided by integrating classical networks such as the current Internet with the Quantum Internet. However, Currently, there is no notion of a "Quantum Packet," - a photonic quantum state along with appropriate headers that function as a single data unit that traverses the quantum network. As no such network stack presently exists for a quantum Internet, this represents a completely unexplored open problem, and its solution requires a multidisciplinary effort, spanning the breath from communications theory and engineering communities to the networking engineering one \cite{Wehner2018vision}.

\section{Concluding Remarks}
The potential to completely change markets and industries - such as commerce, intelligence, military affairs \cite{Cacciapuoti_2020}, tackling classes of problems that choke conventional machines, such as molecular and chemical reaction simulations, optimization in manufacturing and supply chains, financial modeling, machine learning, and enhanced security, has led tech giants, like IBM, Google, Intel, Alibaba, and others to a race for building quantum computers. The Quantum Internet has been proposed as the key strategy to significantly scale up the number of qubits for long-distance communication of quantum and classical information. However, quantum computing and networking technologies are still at an early stage of research and development (R\&D). This section presents the general conclusions on the topic addressed in the text, as well as a summary of the main contributions of the chapter.

Quantum communication networks are a nascent technology. As a society, we have come to expect more out of computing, especially networking, and cannot imagine a day without it. We want networks that are compatible with our computing needs, faster, secure, and accessible anywhere and anytime. The emergence of a quantum-computing paradigm that is radically different and incompatible with the current model of communications presents a unique challenge. Quantum networks have a solid and well-documented quantum-mechanical theoretical foundation but not much is known about translating it into practical implementation for most of the science applications that we presented in this chapter.

As we presented in Section \ref{q_comm_net}, from a communication engineering perspective, the design of the Quantum Internet is not an easy task at all. In fact, it is governed by the laws of quantum mechanics, thus phenomena with no counterpart in classical networks - such as no-cloning, quantum measurement, entanglement and teleporting - would impose terrific constraints on the network design. For instance, classical network functionalities, such as error-control mechanisms (e.g., ARQ) or overhead-control strategies (e.g., caching), are based on the assumption that classical information can be safely read and copied. But this assumption does not hold in a quantum network. As a consequence, the design of a quantum network requires a major paradigm shift to harness the key peculiarities of quantum information transmission, i.e., entanglement and teleportation.

As the size of quantum systems grows, in terms of number of qubits in the case of quantum computers, or physical size/spatial separation in the case of quantum networks, so do the challenges related to connecting different parts of the system while maintaining quantum entanglement across it. For example, long-range communication networks rely on establishing, distributing and maintaining entanglement across  thousands of kilometers. This is challenging due to unavoidable signal losses in the communication channels. As we mentioned in Section \ref{q_repeat}, depending on the tools used for suppressing the imperfections, the quantum information community has identified the following three generations of quantum repeaters: The first generation uses heralded entanglement generation and heralded entanglement purification, which can tolerate more errors but requires two-way classical signaling over the entire chain of quantum repeaters; such signaling then implies that the requisite quantum memory lifetimes/coherence times must be substantially longer than the round-trip communication times. The second generation introduces quantum encoding and classical error correction to replace the entanglement purification with classical error correction, handling all operational errors, which is more demanding in physical resources but requires only two-way classical signaling between neighboring repeater stations, and consequently further improves the quantum communication rate. The third generation of quantum repeaters would use quantum encoding to deterministically correct both photon losses and operation errors. By entirely eliminating two-way classical signaling, the third generation of QRs would promise extremely high entanglement distribution rates that can be close to classical communication rates, limited only by the speed of local operations, in turn, limited by, e.g., photon source rates, detector saturation rates, and timing jitter, etc. 

Another key aspect of a fully functioning quantum Internet is the potential for unconditional information security - a feature of using quantum information that is not possible with classical information processing. A further benefit of using quantum secured information will be that the lifetime of the security is "infinite"; it will be secure against any advances in computation capability that may occur in the future. There have been many cryptographic tasks in which quantum-secured versions have been conceived. For all of these tasks, quantum interconnects are required because of the need to preserve entangled quantum states.

As we mentioned in this chapter, to realize fully the potential of a quantum internet, significant convergent work is still needed to improve the physical hardware. Theoretical work is also required to develop efficient information processing techniques to preserve  quantum information and determine the most robust and secure network connectivity. The development of quantum-secured devices and protocols could transform the cryptographic landscape.

Quantum networks, like many other innovations, which originate from basic research in academia and national labs, face technology transfer challenges despite their overwhelming potential to increase the nation's capabilities and benefit its society. Even though HPC and high-performance optical networks have been developed in concert to improve computing capability, major information technology providers investing in quantum computing, such as Google, IBM, and Microsoft, and others in the telecommunications sector still view quantum networks as a high-risk effort. This means that the government must play an active role in prioritizing and matching investments from the private sector. As a neutral actor, the government could also facilitate the development of standards that will be critical in building inter-operable subsystems critical for quantum telecommunications.

Attentive to this issue, the main countries in the world have been defining as strategic vision focuses on R\&D efforts to advance the development of foundations for the quantum Internet \cite{white_h2020, canada2020, Quantum_init2020}. USA's strategy, for example, was developed through the National Quantum Initiative Act (NQIA), the National Quantum Coordination Office  (NQCO)  and  the  National  Science  and  Technology  Council's  Subcommittee  on  Quantum Information  Science  (SCQIS) and  reflects  deep  community  input  from SCQIS  request  for  information responses of 2018-2019 and from recent workshops hosted by Federal agencies \cite{awschalom2019development}. Quantum Internet Alliance aims to develop a Blueprint for a pan-European entanglement-based Quantum Internet, by developing, integrating and demonstrating all the functional hardware and software subsystems. China already has completed its own quantum key network using satellite communication. The first one, known as Micius, in English, was launched in 2016 \cite{PhysRevLett.120.030501}.

Lastly, quantum networking is a nascent interdisciplinary field in quantum information processing. It is drawing interest from disparate fields such as quantum physics, telecommunications engineering, optical communications, computer science, cybersecurity, and domain science that have not traditionally worked together. Such collaboration is needed to solve a problem as complex as developing a general-purpose quantum network. However, these communities do not currently have a shared vocabulary or world-view, and many do not understand the specifics clients' requirements for interconnecting quantum computers and/or quantum sensors. Therefore, this effort will require a relatively long period of collaboration between researchers from these various communities, so that they can achieve a shared understanding of the problems and of the potential solutions. Once this shared understanding is achieved, it will be possible to perform a more detailed analysis of alternative concepts to determine whiches quantum network architectures will better satisfy the clients' needs.
\newpage

\section*{Acknowledgments}
This research was supported by the Coordena\c{c}\~{a}o de Aperfei\c{c}oamento de Pessoal de N\'{i}vel Superior - Brasil (CAPES) - Finance Code 001, and by the RNP-NSF joint call for research and development in cybersecurity, trough INSaNE (Improving Network Security at the Network Edge) project, funded by the National Science Foundation and the Brazilian Ministry of Science, Technology, Innovation, and Communication (MCTIC) through RNP and CTIC. This work was supported in part by the National Science Foundation under grant CNS-1617437.

\newpage

\bibliographystyle{sbc}
\bibliography{sbc-template}

\end{document}